\documentclass[10pt]{article}
\usepackage{latexsym}
\usepackage{graphics}

\begin{document}

\title{Renormalization group and Ward identities in quantum liquid phases and 
in unconventional critical phenomena}

\author{C. Di Castro$^1$, R. Raimondi$^2$, and S. Caprara$^1$ \\
$^1$INFM-SMC and Dipartimento di Fisica, \\ Universit\`a di Roma 
``La Sapienza'',\\ Piazzale Aldo Moro, 2 -- 00185 Roma, Italy\\
$^2$INFM-NEST and Dipartimento di Fisica, \\ Universit\`a di Roma Tre,\\
via della Vasca Navale, 84 --  00146 Roma, Italy}
\maketitle
\begin{abstract}
By reviewing the application of the renormalization group to different 
theoretical problems, we emphasize the role played by the general symmetry 
properties in identifying the relevant running variables describing the 
behavior of a given physical system. In particular, we show how the 
constraints due to the Ward identities, which implement the conservation laws 
associated with the various symmetries, help to minimize the number of 
independent running variables. This use of the Ward identities is examined 
both in the case of a stable phase and of a critical phenomenon. In the first 
case we consider the problems of interacting fermions and bosons. In one 
dimension general and specific Ward identities are sufficient to show the 
non-Fermi-liquid character of the interacting fermion system, and also allow 
to describe the crossover to a Fermi liquid above one dimension. This 
crossover is examined both in the absence and presence of singular 
interaction. On the other hand, in the case of interacting bosons in the 
superfluid phase, the implementation of the Ward identities provides the 
asymptotically exact description of the acoustic low-energy excitation 
spectrum, and clarifies the subtle mechanism of how this is realized below and 
above three dimensions. As a critical phenomenon, we discuss the 
disorder-driven metal-insulator transition in a disordered interacting Fermi 
system. In this case, through the use of Ward identities, one is able to 
associate all the disorder effects to renormalizations of the Landau 
parameters. As a consequence, the occurrence of a metal-insulator transition is
described as a critical breakdown of a Fermi liquid. 
\end{abstract}
\newpage
\tableofcontents 

\section{Introduction}
\label{intro}
There exist several physical problems where the use of simple perturbative 
methods seems to be deemed to failure from the start. In these cases, already 
in the leading perturbative correction, one faces singular terms which render 
the perturbative expansion meaningless, no matter how small the initial 
expansion parameter is. The great success of the renormalization group (RG) 
has been to turn, at least in a number of highly non-trivial cases, this 
apparent failure into a powerful tool. The RG conceptual scheme \cite{bogoshi,wilson}
teaches us how
singular terms, arising in the perturbative expansion, give rise to the
power-law and scaling behaviors characteristic of the critical phenomena 
\cite{fermili,wilko,revsrg,ma}.  This is  the most successful use of 
the RG, which we only briefly recall in Sec. \ref{hist}. In particular, we introduce the 
basic elements of the RG approach such as universality, scaling, relevant and 
irrelevant variables, and  discuss the Wilson\cite{wilson} and the field-theoretic
RG\cite{CDCJL,revsrg,larkin1969}, which can be seen as two different implementations 
of universality.

The origin of a singular perturbation theory is understood within the 
$\varphi^4$ model for critical phenomena, where $\varphi({\bf r})$ is the 
field whose thermal average specifies the order parameter $\varphi_0$
\cite{landau37}. Indeed, the dimensionless effective coupling constant which 
appears in perturbation theory in $d$ spatial dimensions is 
$u/|t|^{\epsilon/2}$, where $\epsilon=4-d$, $u$ is the coupling constant 
describing the self-interaction of the fluctuations of the field $\varphi$, 
and $t=T-T_c$ measures the deviation of the temperature $T$ from the critical 
temperature $T_c$. This leads to a divergence when approaching criticality 
($t\to 0$), for $\epsilon>0$ ($d<4$) [for the corresponding power counting, 
leading to the expression of the dimensionless coupling constant given above, 
see Fig. \ref{diagr}(a)]. The RG sums the leading singularities into power 
laws for the order parameter $\varphi_0\sim |t|^\beta$, for the susceptibility
$\chi\sim |t|^{-\gamma}$, which measures the linear response to its conjugate 
external field $h$, for the specific heat $C\sim |t|^{-\alpha}$, etc. Here, 
$\beta,\gamma,\alpha$... are called the critical exponents, or critical 
indices.

We will also turn our attention to different uses of the RG approach. One 
class of such applications is met when the critical behavior is 
{\sl unconventional}. This may occur, for instance, when it is difficult to 
identify the order parameter (which can be a complicated object, or a composite
operator) and the related broken symmetry. The problem of the metal-insulator 
transition (MIT) in a disordered interacting electron system is an important 
example of this class
\cite{altshuler1982,nagaoka1985,castellani1985,ando1987,finkelstein1990,belitz1994,hamburg1999}.
In this case the symmetries of the physical problem may provide the guiding 
framework to identify the effective theory. These symmetries can be translated 
into Ward identities which establish relations among the various terms of the 
skeleton structure of the problem, simplifying the RG treatment.

The use of the RG, however, is not limited to the critical behavior. It turns 
out to be extremely useful also in the stable phases. Indeed, there are cases
in which a divergent perturbation theory is the result of relevant low-lying 
excitations in low-dimensional systems, despite the fact that the system is in
a stable liquid phase of the matter. This is the case, e.g., of the Luttinger 
liquid in one-dimensional interacting Fermi systems \cite{solyom} [see Fig. 
\ref{diagr}(b)], or of Bose liquids in the superfluid phase for $d\le 3$ 
\cite{beliaev,gavoret} [see Fig. \ref{diagr}(c),(d)]. The very condition of 
phase stability implies that the susceptibilities (e.g., specific heat, spin 
susceptibility, compressibility, etc.) are finite, and the exact cancellation 
of the singular contributions is controlled by Ward identities. Nonetheless, 
as we shall show, a track of the singularities is usually left  in the 
power-law behavior of some one-particle physical quantities.

In the following, we will touch upon all these examples, often stressing the
key role played in practice by the Ward identities which implement the
fundamental symmetries of the physical system at hand, and for the last two 
cases allow for a complete solution of the low-lying excitation problem.
The rest of the paper is organized as follows. In Sec. \ref{hist}, we
recall a few basic concepts about the RG and how Ward identities may be used
to carry out the renormalization program. In Sec. \ref{rgsp}, we discuss the
application of RG to stable phases. First, in Sec. \ref{llfl}, we examine the 
case of interacting fermions in one \cite{lark,dcm} and low 
dimensions \cite{castellani_94,metzner_98,bares_93,maccarone}. 
Then, in Sec. \ref{ibld}, we analyze the problem of interacting bosons 
\cite{pistolesi,benfatto}. In Sec. \ref{mit}, we focus our attention on the problem of 
disordered interacting Fermi systems
\cite{finkelstein1983,altshuler1983b,altshuler1983,castellani1984,
castellani1984b,castellani1984c,finkelstein1984,finkelstein1984b,castellani1986,
castellani1986b,castellani_98,nussi}.

\section{The renormalization group and the Ward identities}
\label{hist}
\subsection{General remarks on the group transformation}
The first aspect that must be appreciated by looking at the historical
development of the RG is the change of perspective with respect to the 
reduction scheme which, before its introduction, was at work  in formulating a 
condensed-matter theory. The fifties and the sixties  of the past century had 
seen the flourishing of the quasiparticle concept. According to this, an 
interacting system may be effectively described in terms of ``almost'' 
independent elementary excitations (quasiparticles). Implicit in this attitude 
is the privileged role reserved to the ``approximate'' solution of the 
dynamical problem and to the development of the many-body theory, with respect 
to the statistical aspects. In the late sixties and early seventies, the 
realization of the importance of the large-scale fluctuations in approaching 
the critical point shifted the attention onto the statistical problem 
\cite{patapo,kada}. The divergence of the correlation length 
$\xi\sim |t|^{-\nu}$, when $t\to 0$, implies the strong correlation of a large 
number of degrees of freedom within a region of an increasing linear size 
$\xi$, and then invalidates the previous quasiparticle-based reduction schemes.
The collective critical phenomenon does not arise as a simple superposition of 
single microscopic events. Through the action of ordering forces, the 
probability distribution of the collective variables no longer obeys the 
central limit theorem, according to which the correct normalization for an 
extensive variable is provided by the square root of the number of degrees of 
freedom. This violation, via an anomalous normalization of block variables, is 
the mathematical manifestation of the existence of correlations at all length 
scales, so that the crucial hypothesis of statistical independence becomes 
progressively not applicable getting closer to the critical point.

The very fact that $\xi\to\infty$ when approaching criticality, while
invalidating the previous reduction schemes, provides the crucial key for the 
new reduction scheme, i.e., universality \cite{patapo,kada}. 

There are two equivalent forms of universality. First of all, the microscopic 
details which specify the peculiarities of each individual system become 
gradually irrelevant when approaching criticality, and the infinitely large 
number of degrees of freedom involved is well accounted for by a small set of 
relevant variables. Once the proper choice of the relevant variables is made 
(e.g., $t$ and $h$ or $\varphi$), one can change the parameters $\zeta_i$ 
which specify the microscopic details (unless they assume a value which 
changes the symmetry of the problem, thus becoming relevant), without changing 
the critical behavior of the system. Said in other words, systems which 
differ in the irrelevant variables share the same critical behavior 
\cite{kadanoff0}. This condition translates into the invariance of 
the (singular part of the) free energy ${\cal F}$, or equivalently of its 
Legendre transform $\Gamma=\int d^d {\bf r}~h\varphi-{\cal F}$,
\begin{eqnarray*}
{\cal F}(t,h;\{\zeta_i\})&=&{\cal F}(t',h';\{\zeta_i'\}),\nonumber\\
\Gamma(t,\varphi;\{\zeta_i\})&=&\Gamma(t',\varphi';\{\zeta_i'\}),\nonumber
\end{eqnarray*}
i.e., a suitable rescaling of the relevant variables fully accounts for
a change in the irrelevant variables.

The second form of universality relies on the fact that the unit length scale 
is irrelevant when $\xi\to\infty$. This is translated into the statement that 
one can eliminate the degrees of freedom at short distance, by grouping
microscopic variables into ``block variables'', 
within an iterated procedure to build larger and larger blocks, without 
changing the critical behavior of the system \cite{kada}.

The field-theoretic RG equations generalize the universality relations in the 
sense that they relate one model system to another by varying the coupling 
constant $\zeta$ and suitably rescaling the other variables and the 
correlation functions \cite{CDCJL}. In ordinary critical phenomena, described 
by the $\varphi^4$ theory, $\varphi$ (wave-function) and $t$ (mass) must be 
rescaled. When the coupling reaches its fixed point, a two-parameter scaling 
follows.

The Wilson new RG implements the second idea of universality illustrated 
above, by eliminating the short-wavelength fluctuations of the field 
$\varphi$, i.e., those with characteristic momenta $\Lambda/s<q<\Lambda$, 
where $\Lambda$ is the ultraviolet cutoff, and $s$ is the scaling parameter
\cite{wilson}. The real space version of the Wilson RG implementation is the bloc
variable transformation\cite{dcjl,jona1974,cassandro1975,joca}.

The calculation of the critical indices was performed either numerically, 
through recurrence equations, or analytically, within the $\epsilon$ expansion 
\cite{wilsonfisher}. This latter procedure relies on the perturbative 
expansion of the RG transformation in the parameter $\epsilon=4-d$, starting 
from the known case $d=4$. The direct calculation of the divergent critical
quantities is thus avoided.

In general, a sets of parameters $\{\mu_j\}$ specifies a given Hamiltonian $H$
in the space ${\cal H}$ of the Hamiltonians. The action of the RG 
transformation ${\cal R}_s$ (which depends on the scaling parameter $s$) in 
this space preserves the functional $\Gamma$, so that symbolically 
\cite{jona1974,cassandro1975}:
$$\Gamma[{\cal R}_s[H]]=\Gamma[H].$$
The above equation can be written in a differential form as
$$d_{{\cal R}_s} \Gamma=0.$$
The functional derivatives of $\Gamma$ yield the group equations for all the 
physical quantities.

A fixed point of the transformation is such that $H^*={\cal R}_s[H^*]$.
At criticality $\xi=\infty$ and it remains infinite under iteration 
of RG transformations. By iteration of the group transformation on the critical
surface one model Hamiltonian is transformed into another, until a fixed-point
simplified Hamiltonian is reached as $s\to\infty$, which is manifestly scale
invariant, and all the irrelevant transient terms have been eliminated.
The relevant parameters define a finite set of relevant directions of escape
from a given fixed point \cite{kadaweg}, and their scale dependence gives
a microscopic definition of the critical indices. Universality corresponds to 
the domain of attraction of a given fixed point. Within a given domain, the 
transformations become asymptotically equivalent\cite{dcjl,jona1974}.

Standard renormalizability implies that the removal of the ultraviolet cutoff, 
$\Lambda\sim 1/a \to \infty$ (where $a$ is the lattice spacing or some other
microscopic characteristic length scale) only requires the definition of a 
finite (small) set of renormalized parameters. Relevant parameters and 
renormalized parameters coincide asymptotically in the infrared region.

Here a general problem arises. Recently, a renewed attention has been devoted 
to the exact RG equations \cite{wegner2001}. In practice, however, one has to 
truncate these equations or reduce the number of flow parameters. In this last 
case, one  has to initialize the action to follow a renormalized trajectory 
with a small number of flowing parameters, and an infinite set of parameters 
in the action must be specified. This specification is a highly 
non-perturbative condition.

In practice, therefore, one should rather choose the best simple action, 
guided by the symmetries of the problem and by the physical conditions. The 
knowledge of the fundamental symmetries inherent to each specific problem, and 
in particular those of the order parameter, have to be assumed in order to 
make the proper choice of the basic variables on which the RG transformation 
acts.

\subsection{Various exploitations of the Ward identities}
The RG approach takes enormous advantage from the explicit implementation
of the symmetries of the original problem, which translate into Ward identities
relating  various independent quantities.

First of all, the Ward identities allow for a reduction of the renormalization
parameters.  The case of quantum electrodynamics is a textbook example. The 
local $U(1)$ gauge invariance of the Lagrangian, which is invariant under the 
simultaneous transformation of the phonon field 
$A_\nu\to A_\nu+\partial_\nu\vartheta$ and of the fermion field 
$\psi\to \psi {\rm e}^{ie\vartheta}\simeq (1+ie\vartheta)\psi$, where 
$\vartheta$ is the gauge field, allows to relate the self-energy and vertex 
renormalizations, whose corrections correspond to superficially divergent 
diagrams within perturbation theory in $d=3+1$ dimensions.

Secondly, the Ward identities provide the guiding framework for the
identification of the proper running variables (effective coupling constants),
and how they are related to physical quantities. We will consider the case of 
disordered interacting electron systems. As we shall see, the various
interaction amplitudes are dressed by disorder. The renormalization parameters
of the corresponding field-theoretical formulation of the problem (the
non-linear $\sigma$-model \cite{finkelstein1983,finkelstein1990}) are identified in terms of 
physical quantities of the Fermi-liquid theory by requiring the gauge 
invariance of the skeleton structure of the response functions
\cite{altshuler1982,castellani1984,castellani1984b,castellani1986,castellani1986b}. 

Finally, the Ward identities control the exact cancellation of the 
singularities in the various response functions within stable phases, despite
a singular perturbation theory. In these cases, indeed, specific symmetries 
are related to additional Ward identities, which enforce the cancellations.

The combination of the three procedures indicated above leads to the
asymptotically exact description of: the Luttinger liquid in $d=1$ 
\cite{solyom,lark}, and its crossover to the Fermi liquid in $d>1$ 
\cite{castellani_94,metzner_98}; the non-Fermi-liquid behavior of an 
interacting electron system with singular forward scattering in $d>1$ 
\cite{bares_93,maccarone}; the Bose liquid in the presence of 
the Bose-Einstein condensate \cite{pistolesi}; the disordered interacting 
electron systems \cite{finkelstein1983,castellani1984}. In Tab.
\ref{table1} we provide a scheme of the symmetries and Ward identities
which are relevant in the various physical problems discussed in this paper.

Each specific system has been discussed at length in several papers. We 
emphasize here the general common aspects and give a unifying view of these
apparently very different theoretical problems.

\section{Renormalization group and  stable phases}
\label{rgsp}
As we have already indicated in Sec. \ref{intro}, there are cases where the
perturbation theory is singular even in a stable phase. We first recall here 
the case of the Luttinger liquid in one dimension, its dimensional crossover 
to a Fermi liquid as soon as $d>1$ in the presence of short range forces, and 
the non-Fermi-liquid behavior when the forces among the fermions are 
sufficiently singular. We then consider the low-lying excitations from a 
ground state of bosons in the presence of interaction, where the condensate 
changes the power counting and leads to singular terms in perturbation theory 
for $1<d\le 3$. The use of the RG, in these cases, acquires a different 
meaning with respect to critical phenomena, as the thermodynamic stability
implies here a cancellation of singularities at any order in perturbation 
theory in all the susceptibilities, like specific heat, compressibility, etc. 
Behind these exact cancellations there must be specific symmetry properties, 
which can be used in the form of Ward identities to close the hierarchy of 
RG equations and solve the problem. Although a stable system is characterized
by finite response functions, the re-summation of singularities may still manifest 
itself with an anomalous dimension as in the one-particle fermion Green's function.

\subsection{Interacting fermions in low dimensions}
\label{llfl}
Within a RG approach, the effective low-energy Hamiltonian which describes 
a metallic system can be obtained via the Wilson-like iterative elimination of 
states outside a shell of thickness $\Lambda$ around the Fermi surface 
\cite{solyom,metzner_98,gallavotti}. As $\Lambda$ is iteratively 
scaled to zero a fixed point is reached in $d=3$ which corresponds to a gas of 
quasiparticles with finite residual Hartree-like interaction, described by the 
Landau functional. Ordinary three-dimensional metals are thus well described 
by the Landau Fermi-liquid theory \cite{nozieres}. The interaction is 
effectively taken into account in the various response function through a 
small set of Landau parameters, entering the physical quantities, e.g.,
the specific-heat coefficient $\gamma=C_V/T$, 
the spin susceptibility $\chi$, the compressibility $\partial n/\partial\mu$, 
and the spectral weight of the Drude peak in the optical conductivity, all of them 
being  finite.

A finite wave-function renormalization $Z<1$ reduces the discontinuity of the 
occupation number in momentum space at the Fermi surface with respect to the 
value of the Fermi gas ($Z=1$). The elementary excitations of the system are 
the single-particle excitations (quasiparticles), which carry both charge and 
spin, and the collective charge and spin modes, which get overdamped when they 
enter the particle-hole continuum of excitations.

In recent times, the interest for non-Fermi-liquid metallic phases has been 
rekindled by the observation of anomalous features in single-particle and
transport properties of the cuprates, which have been interpreted as a 
signature of non-Fermi-liquid behavior \cite{anderson_book}. These compounds 
are layered (quasi-two-dimensional) materials, which are insulating when 
stoichiometric, and become metallic upon chemical doping. The metallic phase 
is strongly anomalous at low doping and gradually (and smoothly) evolves 
towards a Fermi-liquid-like metal at large doping.

The breakdown of the Fermi liquid could be produced by the suppression of
quasiparticle spectral weight at the Fermi surface due to the generation of an 
interaction-induced anomalous dimension in the wave-function renormalization, 
$Z\sim |p-p_F|^\eta$, where $p-p_F$ measures the deviation of the quasiparticle
momentum $p$ from the Fermi momentum $p_F$. In such a case, $Z$ vanishes as
$p\to p_F$, the low-lying single-particle excitations are completely 
suppressed, and the low-energy behavior of the system is dominated by the 
charge and spin collective modes. Since these have, in general, different 
velocities, the system is characterized by the so-called charge and spin 
separation.

This behavior is achieved in $d=1$, where the metallic phase is a Luttinger
liquid. The hint for the breakdown of the Fermi-liquid picture is given by the 
appearance of logarithmic singularities in perturbation theory as $p\to p_F$ 
[see Fig. \ref{diagr}(b)]. These singularities are controlled by the RG, and
an effective field theory results, with marginal forward scattering, the
so-called Luttinger model, which is exactly solvable. Additional conservation 
laws, which are peculiar to one-dimensional systems with forward scattering, 
constrain the model and lead to a closure of the equations of motion.

Indeed, the total charge and spin conservation is translated into the usual 
Ward identity
\begin{equation}
\omega\Lambda_0-{\bf q}\cdot{\mathbf \Lambda}=
G^{-1}(\varepsilon+\omega,{\bf p}+{\bf q})-G^{-1}(\varepsilon,{\bf q}),
\label{usualwi}
\end{equation}
which connects the irreducible density ($\Lambda_0$) and current
(${\mathbf \Lambda}$) vertices to the electronic Green's function $G$. In 
addition, in this case, charge and spin are separately conserved at each Fermi 
point $p=\pm p_F$, leading to an additional Ward identity \cite{lark}
stating the proportionality of the current vertex to the density vertex, due to
the unique momentum direction,
\begin{equation}
{\mathbf \Lambda}=v_F {\hat{\bf p}}\Lambda_0,
\label{newwi}
\end{equation}
where $v_F$ is the Fermi velocity and the direction versor ${\hat {\bf p}}$ 
here ($d=1)$ is $\pm 1$. By means of the two Ward identities (\ref{usualwi}) 
and (\ref{newwi}) one is able to express the density vertex $\Lambda_0$ as a 
function of the fermion propagator $G$, thus closing the Dyson equation of 
Fig. \ref{dysonfig}.

The effective interaction $D$ (which is represented by the tick-dashed line in Fig.
\ref{dysonfig})  contains the charge and spin collective modes via
the RPA re-summation of the bare interaction with the fermion polarization 
bubble. The RPA re-summation is shown to be exact by the use of the same Ward 
identities (\ref{usualwi}) and (\ref{newwi}). As a result density and spin 
response functions are finite, leading to a stable metallic phase.

On the contrary, from the exact evaluation of $G$ one finds that its behavior 
is controlled by a non-universal anomalous dimension $\eta$. Single particles 
hence move as composite objects due to the strong mixing with the charge and 
spin collective modes, induced by the effective interaction.

The question then arises whether the Luttinger-liquid behavior can be extended
to higher dimensions. The analysis of the dimensional crossover shows that 
this is not possible if the interactions are short-ranged. Indeed, the generic 
integrals in $d$ dimensions
$$
\int d^d{\bf q}~f({\bf q},{\bf p}\cdot{\bf q})=
S_{d-1}\int dq ~q^{d-1}\int_0^\pi d\theta (\sin \theta)^{d-2}
f(q,pq\cos\theta),
$$
where $S_{d-1}$ is the solid angle in $d-1$ dimensions, are strongly peaked at 
$\theta=0,\pi$ for $1\le d <2$, so that all the relevant vectors are still 
parallel or antiparallel. In this case, the additional Ward 
identity (\ref{newwi}) is still valid asymptotically near the Fermi surface.
However, the mixing with the collective modes is reduced as soon as $d>1$, 
since the effective interaction $D({\bf q},\omega)$ is averaged over the $d-1$ 
momentum components perpendicular to the Fermi momentum. The effective 
interaction, which is marginal in $d=1$, scales to zero for $d>1$, and the 
system is a Fermi liquid.

However, if we consider a long-range bare interaction among the fermions,
$V(q)\sim q^{-\alpha}$, the effective interaction dressed by the RPA series is
\begin{equation}
D({\bf q},\omega)=\frac{1}{q^\alpha}~
\frac{\omega^2}{\omega^2-c^2q^{2-\alpha}},
\label{efflr}
\end{equation}
and is dominated by a collective mode $\omega(q)=cq^{1-\alpha/2}$, which is 
propagating and gapless for $\alpha<2$ (the parameter $c$ depends on the 
microscopic parameters of the fermion model, and does not acquire 
singular corrections, see below). This singular behavior as $q\rightarrow 0$ 
compensates the rescaling to zero of the effective interaction due to its 
averaging over the transverse momentum and leads to a non-Fermi-liquid 
behavior for $\alpha\ge 2d-2$ \cite{bares_93}.

A singular interaction may appear in a system close to an instability, due to
the coupling of the fermion quasiparticles with the critical fluctuations.
Various proposals for the breakdown of the Fermi-liquid picture in the cuprates
rely on the existence of a quantum critical point in their phase
diagram \cite{qcp}.

The problem of fermions interacting via a singular interaction is plagued by 
infrared divergences and requires a proper RG treatment, supported by the Ward 
identities \cite{maccarone}.

The vertex $\Lambda_0$ induces a coupling constant $g$ which in principle
contains the divergences of the vertex (which requires the renormalization 
$Z_V$), of the fermion propagator (renormalized by $Z$) and of the 
collective-mode propagator $D$ (renormalized by $Z_D$). The coupling is 
accordingly renormalized as $g_0=g Z Z_D^{1/2}/Z_V$. However, the total charge 
conservation translated into the usual Ward identity (\ref{usualwi}),
gives, in the dynamic limit,
\begin{equation}
\label{zvz}
Z_V^{-1}=\lim_{\omega\to 0}\lim_{{\bf q}\to 0}\Lambda_0=
\partial_\varepsilon G^{-1}\equiv Z^{-1}.
\end{equation}
Moreover, since small-${\bf q}$ scattering dominates, the charge is 
asymptotically conserved at each point of the Fermi surface, leading to the 
additional Ward identity (\ref{newwi}). When the vertex $\Lambda_0$ is 
accordingly expressed in terms of the fermion propagator $G$, one realizes 
that the effective interaction $D$, i.e., the collective-mode propagator 
(\ref{efflr}) does not get anomalous 
corrections beyond the RPA approximation, i.e., $Z_D=1$. 
In fact, the vertex corrections to the bare polarization bubble 
asymptotically cancel with the selfenergy corrections, thank to the
additional Ward identity (\ref{newwi}).

Therefore, together 
with Eq. (\ref{zvz}), this implies that $g=g_0$, and the dimensionless
running coupling 
constant $u=g_0^2/s^{x_u}$, where $s$ is some infrared cutoff which plays the
role of the scaling parameter, evolves under RG with its bare dimension 
$x_u=1-d+\alpha/2$ as $s\to 0$. Therefore at the dimension 
$d=d_c\equiv 1+\alpha/2$ the coupling is marginal ($x_u=0$). For $d<d_c$ the 
bare dimension $x_u$ is positive,  the theory scales to strong coupling and 
the Fermi liquid breaks down. Finally, for $d>d_c$, $x_u$ is negative and the 
effective coupling scales to zero. Thus $Z$ stays finite and the system is a 
Fermi liquid.

From the one-loop perturbative RG, one finds the explicit form of the 
wave-function renormalization as a function, e.g., of the quasiparticle
energy $\varepsilon$. At $d=d_c$,  $Z\sim|\varepsilon|^{u_0}$, vanishes 
with a non-universal exponent $u_0\sim g_0^2$, as in the Luttinger liquid. For 
$d<d_c$, instead, $Z$ scales to zero as a stretched exponential, for 
$\varepsilon\to 0$. In both cases, the single-particle spectral weight is 
suppressed as the Fermi surface is approached.

\subsection{Interacting bosons in low dimensions}
\label{ibld}
The theory of the interacting Bose system has been motivated by the observation
of superfluidity in liquid helium for $T<T_\lambda=2.17$K. The connection 
between the form of the spectrum of the low-lying excitations and the 
superfluid properties of the system was established by Landau via the 
criterion which requires a finite minimum slope of the quasiparticle 
dispersion $\varepsilon ({\bf q})$ \cite{agd_bos}.

The simplest interacting Bose system assumes a quartic contact interaction $v$ 
for the boson field $\varphi$. The mean-field solution due to 
Bogoljubov \cite{bogol},  is obtained by linearizing the interaction term via 
the factorization of the condensate density 
$n_0^{1/2}=\langle\varphi\rangle=\langle\varphi^\dagger\rangle$. The 
free-particle spectrum $q^2/2m$ is converted at low momenta into a linear 
spectrum $\varepsilon ({\bf q}) =\sqrt{2 n_0 v}q\equiv c_0 q$, characteristic 
of a sound-like excitation in the liquid, leading to superfluidity.

However, the Bogoljubov solution leads to a non-zero anomalous self-energy
$\bar\Sigma=v\langle \varphi\varphi\rangle=vn_0$. Moreover, the first 
corrections to the Bogoljubov approximate solution, within the so-called 
pairing approximation, lead to a spurious gap in the excitation spectrum
\cite{ga}. Both these results are in contrast with exact results, which show 
that no gap is present in the spectrum (Hugenoltz-Pines theorem) \cite{hp}, 
and that the anomalous self-energy vanishes, $\bar\Sigma\equiv 0$ \cite{NN}.

On the other hand, standard perturbation theory is plagued by infrared 
divergences in $d\le 3$, due to the Goldstone sound mode 
\cite{beliaev,gavoret}, despite the fact that the superfluid phase is a stable 
liquid phase of matter. All these aspects call for a cautious analysis of the 
problem, as it was first recognized in Ref. \cite{gavoret}. Benfatto used
the Wilson RG approach to determine the scaling behavior for a number of running
variables \cite{benfatto}. However, his treatment was limited to $d=3$ and did
not take advantage of a systematic implementation of the Ward identities. We
summarize the results in generic $d$ dimensions, fully exploiting the Ward
identities, along the lines of Ref. \cite{pistolesi}.

Within the standard formulation of the problem, an external source
$A_\nu=(\mu ({\bf r}),{\bf A}({\bf r}))$, $\nu=0,...,d$, coupled to the boson 
density (current) and a field $h({\bf r})$ conjugate to $\varphi({\bf r})$ are 
introduced to generate the correlation functions. The original problem is 
recovered when ${\bf A}\to 0$, $h\to 0$ and $\mu$, taken as a constant, is 
identified with the chemical potential of the liquid. By introducing the 
longitudinal and transverse components to the direction along which the 
symmetry is broken, we have $\varphi =\varphi_L+{\rm i}\varphi_T$, with 
$\langle \varphi_L\rangle_{h\to 0}=\sqrt{n_0}$, and $h=h_L+{\rm  i}h_T$. 
Functional derivatives of the free energy ${\cal F} [A_{\nu},h_i]$
with respect to the 
external sources produce the various density, current expectation values and
Green's functions, e.g.,
$\varphi_{i0}({\bf r})\equiv\langle \varphi_i\rangle=
\delta {\cal F}/\delta h_i$, with $i=L,T$.

The Legendre transform $\Gamma[A_\nu,\varphi_{i0}]=\int d^d{\bf r}
\sum_ih_i\varphi_{i0}-{\cal F}[A_\nu,h_i]$ is the generating functional of the 
various vertex functions, e.g., 
$h_i({\bf r})=\delta \Gamma/\delta \varphi_{i0}$.

In this representation the mean-field Green's functions in frequency and 
momentum space read
\begin{equation}
\left\{\matrix{
G_{LL}&=&\displaystyle{\frac{q^2}{\omega^2-c_0^2q^2},}\cr
G_{LT}&=&\displaystyle{\frac{\omega}{\omega^2-c_0^2q^2},}\cr
G_{TT}&=&\displaystyle{\frac{c_0^2}{\omega^2-c_0^2q^2},}}\right.
\label{greenbos}
\end{equation}
whence it is seen that the most singular contribution in the infrared is 
carried by $G_{TT}$. In the presence of the condensate, the four-leg vertex
associated to the coupling constant $v$ yields various four- and three-leg
vertices, associated to the various coupling constants $v_{LLLL}$, etc., 
$v_{LLL}$, etc. (see Fig. \ref{figbos}). Starting from Eq. (\ref{greenbos}),
the dimensional analysis, in units such that $[q]=[\omega]=1$, identifies nine
relevant or marginal running variables with singular corrections in $d\le 3$.

However, the number of independent parameters is reduced by the exploitation
of the local gauge invariance of the model. The free energy ${\cal F}$, or
equivalently  $\Gamma$, are invariant under the local transformation generated 
by the operator
$$
{\cal T}=\left[\begin{array}{cc}
\cos\vartheta &-\sin\vartheta \\
\sin\vartheta &~~ \cos\vartheta
\end{array}\right]\simeq 1+\vartheta
\left[\begin{array}{cc}
0 &-1 \\
1 &~~0
\end{array}\right],
$$
which acts in the $L,T$ space, i.e.,
\begin{equation}
\label{bosesymmetry}
\Gamma[A_\nu+\partial_\nu\vartheta,{\cal T}_{ij}\varphi_{j0}]=
\Gamma[A_\nu,\varphi_{i0}],
\end{equation}
with $i=L,T$ and $j=L,T$.

Functional derivatives of $\Gamma$ with respect to $\vartheta$, $A_\nu$, and $\varphi$ 
generate the various Ward identities. The first two are obtained by a 
functional derivativation   with respect to $\vartheta$ and then either 
with respect to $\varphi_L$ or $\varphi_T$. In this way the Ward identities
connect the two-point vertices to the external fields, showing, e.g., that 
when $h\rightarrow 0$ no gap is present in the spectrum. In this way the 
Hugenholtz-Pines theorem is simply recovered. Three-point vertices
are related 
to two- and four-point vertices via three relations, which are the 
implementation of the continuity equation for this special case in the 
presence of the condensate. By then, all the strongly relevant coupling 
constants are fixed, and one is left with four marginal running variables. 
However, three of them are connected to physical quantities. In fact, they are 
related to the derivatives of the vertices with respect to frequency or 
momentum, and the corresponding Ward identities involve density and current 
correlation functions, thus implying relations to superfluid density, 
condensate compressibility, sound velocity, all of them free of
divergences. Therefore, these identities 
guarantee the exact cancellation of the singular contributions in these 
variables, which are then RG invariants.

After all the identifications are made, one running variable is left, e.g., 
the longitudinal two-point vertex coupling $v_{LL}$.

The one-loop RG equations  reproduce this
situation of being left with only $v_{LL}$ as running coupling, with
$v_{LL}\sim s^{\tilde\epsilon}$, with $\tilde\epsilon=3-d\ge 0$, at 
$d\le 3$. In this case the anomalous self-energy
$\bar\Sigma$ vanishes. 

For $d>3$, instead, $v_{LL}=2\mu$ is finite, and the 
Bogoljubov result is recovered. In both cases, the spectrum is linear, leading 
to sound-like low-energy excitations. However this spectrum is realized in a 
completely different way in $d\le 3$.

As in the Luttinger liquid for the one-particle Green's function, the behavior 
$v_{LL}\sim s^{\tilde\epsilon}$ is fixed exactly by the Ward identities, which 
allow to close the (dominant part of the) equation for $v_{LL}$
(see Fig. \ref{figbos}) since the three-leg vertex 
$v_{TTL}$ is identified with $v_{LL}/\varphi_{L0}$ via Ward identities.
When dealing with RG loop expansions, further corrections to the one-loop RG 
results cannot change the power-law exponents.

\section{An example of unconventional criticality: the metal-insulator transition}
\label{mit}
\subsection{Metal-insulator transition in disordered electron systems}
When we deal with disordered electron systems, the singularities do not come 
directly from integrations of loops in terms of  the original fermion 
propagators, as in Fig. \ref{diagr}, but rather from the soft collective modes 
related to diffusion, in terms of which an effective action can be derived for 
both non-interacting \cite{efetov1980,hikami1981} and interacting systems
\cite{finkelstein1983}. In this case, we will therefore 
enter in rather more details  with respect to the previous topics.

According to basic quantum mechanics, within the one-electron approximation,
insulating and metallic behavior occur as a result of the complete or
partial filling of the highest occupied energy band of a solid, respectively.
Beyond the one-electron approximation, correlation effects may lead to
a drastic rearrangement of the energy levels and give rise to insulating
behavior in situations where the metallic one is expected on the basis of
band filling. This is usually referred to as a Mott phenomenon. 

Scattering from randomly located impurities 
may also lead to a transition from a metallic to an insulating behavior. In 
this second case, rather than a rearrangement of the energy levels, the 
coherent scattering from the impurities modifies the phase of the electron 
wave function in such a way that it may result localized in space, 
thereby yielding an insulating behavior \cite{anderson1958}.  
Experiments were performed both in metallic and semiconducting systems. These 
latter, in particular, turned out to be better suited for the study of the 
MIT, since the degree of disorder could be changed by the doping level, as it 
happens, for instance, in silicon doped with phosphorus (Si:P). 
There exist several review articles, which summarize the development of the
field at various stages
\cite{altshuler1982,nagaoka1985,castellani1985,ando1987,finkelstein1990,belitz1994,hamburg1999}.
In recent years, there has been a renewed interest in the MIT, which is very 
actively investigated in the two-dimensional electron gas of MOSFET and 
heterostructure devices (for a recent review see, e.g., Ref. 
\cite{abrahams2000}).

At theoretical level, a crucial element was the discovery that the
quantum corrections to the classical Drude formula for the electrical
conductivity give rise  to singular terms in low-dimensional systems.
These corrections arise as a result of the quantum interference of
electron waves in disordered systems. There are two types of such terms. The 
first, known as the weak-localization (WL) correction, is a purely
one-particle effect and is due to the interference of time reversed 
trajectories \cite{abrahams1979,gorkov1979} 
(see Fig. 
\ref{weaklocalization_fig}). 
The second is a consequence of the enhancement of the
electron-electron interaction in a disordered system and is usually referred
to as electron-electron interaction (EEI) correction
\cite{altshuler1979,altshuler1980} (see Fig. 
\ref{effectiveconductivitydiagrams}). 

In two dimensions, in particular, both the WL and the EEI correction 
to the conductivity are logarithmic,
\begin{equation}
\label{weaklocalization}
\Delta\sigma_{\rm WL} = 
-{e^2\over \pi^2\hbar } \ln(L_\phi / l ),
\end{equation}

\begin{equation}
\label{interactioncorrection}
\Delta\sigma_{\rm EEI} =
- {e^2 \over \pi^2 \hbar } \ln( L_T / l).
\end{equation}
Both the above effects have been discussed in great detail in the literature 
and we do not provide a derivation of Eqs.
(\ref{weaklocalization},\ref{interactioncorrection}) here, but limit ourselves
to a comment on the EEI correction, after Eq. (\ref{diffuson}).

In Eqs. (\ref{weaklocalization},\ref{interactioncorrection}), the argument of 
the logarithmic corrections contains the ratio of two length scales. The 
first is the mean free path, $l=v_F\tau$, $\tau$ being the elastic scattering 
time, which sets the microscopic scale, beyond which the system behaves 
diffusively. The second length scale differs in the two cases. For WL it is 
$L_{\phi}$, the scale over which inelastic scattering starts to destroy the 
interference effects and one can show that it becomes infinite at zero 
temperature \cite{altshuler1982b}. For EEI it is usually given by the
thermal length $L_T$. In a diffusive system, all length scales correspond to
characteristic times via the diffusion constant $D=v_F^2\tau/2$: the elastic 
scattering time $\tau =l^2 /D$, the
dephasing time $\tau_{\phi}=L^2_{\phi}/D$, and the thermal time
$\tau_T=\hbar /k_B T =L^2_{T}/D$. As a result, the above corrections become
singular at low temperature.

By recalling the Drude formula for the electrical conductivity 
$\sigma =2  e^2N_0 D$, where $N_0$ is the single-particle density of states
at the Fermi level, one sees that the corrections 
(\ref{weaklocalization},\ref{interactioncorrection}) in units of $\sigma$ are 
controlled by the factor $t=1/(4\pi^2N_0D\hbar)$ (not to be confused with the 
deviation from the critical temperature of the previous sections). The 
parameter $t$ is related to the two-dimensional conductance $g=G/(e^2/h)$, by 
$t=1/(g\pi)$. In good metals, normally, $g\gg 1$ and $t$ is a small parameter. 

The WL correction has been used to formulate a scaling theory of the 
MIT
\cite{efetov1980,abrahams1979} (for a pedagogical introduction
see, e.g.,  Nagaoka's and  Kawabata's contributions in Ref. 
\cite{nagaoka1985}).
This theory has a unique scaling variable, the
dimensionless conductance $g$, which according to the
Thouless's argument describes how the eigenstates of the
system change when we add together blocks of size $L$ to make
blocks of size $2L$ \cite{thouless1974}. There are two energy scales to be
considered. The first is the energy level spacing that goes as
$\Delta E =1/(dN/dE)=1/(N_0 L^d)$. The second
is the energy perturbation due to the  change in the boundary conditions.
This may be related to the travel time in a diffusive system of
size $L$, i.e.,  $\delta E\approx \hbar /\tau =\hbar \sqrt{D/L^2}$. The
ratio of these two energies gives
$$
\frac{\delta E}{\Delta E}=\frac{1}{2\pi}\frac{\sigma}{2e^2/h}L^{d-2}
=\frac{g}{2\pi}.
$$
Starting from the metallic side, where $\sigma$ is a constant, we see that in 
two dimensions the conductance 
$g$ becomes marginal. By assuming a scaling behavior, one can derive a RG 
flow equation for $g$ from the WL correction in Eq. (\ref{weaklocalization}).
As a result one predicts that, in two dimensions, the RG flow drives the 
system to an insulating ground state, with $g\rightarrow 0$. The EEI shows
qualitatively the same effect and one wonders whether one has to introduce
further scaling parameters to take into account the interaction strength
\cite{castellani1983}. The answer to such a question is far from trivial and 
has required quite some work. In particular, it has been understood that, 
within a Fermi-liquid description, the Landau parameters provide the 
additional running variables whose RG flow has to be studied together with the 
conductance $g$ \cite{finkelstein1983,castellani1984}. In order to emphasize 
the physical origin of the additional running variables, we show in the next 
section how the interplay of interaction and disorder gives rise to singular 
corrections to thermodynamic quantities. This is to be contrasted with the WL 
correction which leaves the thermodynamic properties unaffected.

\subsection{Thermodynamics}
We present the derivation  of the correction to the
thermodynamic potential due to the interaction and disorder
\cite{altshuler1983,castellani1986,schwab1999}.
 The coupling to the impurities
is described by the Hamiltonian
$$
H_{imp}=\int {\rm d}{\bf r} U({\bf r})
\psi_{\alpha}^{\dagger}({\bf r})\psi_{\alpha}({\bf r}),
$$
while the electron-electron interaction  is of the standard form
$$
H_{int}=\frac{1}{2}\int {\rm d}{\bf r} {\rm d}{\bf r}' V_0({\bf r}-{\bf r}')
\psi_{\alpha}^{\dagger}({\bf r})
\psi_{\beta}^{\dagger}({\bf r}')\psi_{\beta}({\bf r}') \psi_{\alpha}({\bf r}).
$$
Greek indices indicate spin indices and sum over repeated indices is 
understood. To begin with, let us consider the first-order 
exchange-interaction correction to the thermodynamic potential (see Fig. 
\ref{thermo})
\begin{equation}
\label{th2}
\Delta \Omega =-\frac{1}{2} T^2\sum_{\omega ,\varepsilon}
\int {\rm d}{\bf r} {\rm d}{\bf r}'
 V({\bf r}-{\bf r}', \omega)G({\bf r},{\bf r}';\varepsilon)
G({\bf r}',{\bf r};\varepsilon +\omega),
\end{equation}
where we have introduced a retarded electron-electron interaction
$V({\bf x}-{\bf x}',\omega )$ to take into account screening effects  which 
we will come back to later on. $\omega$ and $\varepsilon$ are Matsubara boson and 
fermion frequencies, respectively. Notice that the electron Green's function 
depends separately on its spatial arguments, since translational invariance 
does not hold in the presence of disorder. In general, however, the impurities 
are randomly distributed and it is enough to average over them. For the 
average quantities the translational invariance is restored. The average can 
be performed by means of the well-known impurity technique\cite{agd_bos}. In Eq. (\ref{th2}) 
this implies to average the product of the two Green's functions. We assume a 
Gaussian, $\delta$-correlated impurity potential $U({\bf r})$ with
\begin{equation}
\label{th3}
\langle U({\bf r}) U({\bf r}') \rangle =
{1\over 2 \pi N_0 \tau } \delta( {\bf r} - {\bf r'} ),
\end{equation}
and use a self-consistent Born approximation for the electron self-energy 
(see Fig. \ref{thermo})
\begin{eqnarray}
\label{th4}
 \Sigma( {\bf r},t ; {\bf r}' ,t' )& = &
{1\over 2 \pi N_0 \tau}    G( {\bf r},t; {\bf r}, t' )
\delta({\bf r} - {\bf r}').
\end{eqnarray}
When the above self-energy is used to evaluate the Green's function, one obtains
a self-consistency equation which leads to a finite lifetime $\tau$. The 
average procedure is limited to correlations of impurity insertions according 
to Eq. (\ref{th3}). Insertions  made on the same Green's function are 
automatically taken into account in Eq. (\ref{th4}). When one has an average 
of two or more Green's functions, as in evaluating the response functions,
 correlations between insertions on different 
Green's functions must also be considered. To this end one re-sums the infinite 
series of the so-called {\sl ladder} diagrams (see Fig. \ref{ladder})
$$
L({\bf q}, \omega)=1+\eta +\eta^2+...=\frac{1}{1-\eta},
$$
where
$$
\eta =\frac{1}{2\pi N_0\tau}\sum_{\bf p}
G({\bf p},\varepsilon)
G({\bf p}+{\bf q},\varepsilon +\omega).
$$
The sum over the momentum ${\bf p}$ differs from zero only when the Matsubara
energies $\varepsilon$ and $\varepsilon +\omega$ have opposite signs. For 
vanishing frequency and momentum, $\eta\rightarrow 1$ signaling the emergence 
of a (diffusive) pole in the ladder. Indeed, in the low-frequency and momentum
approximation, $\omega\tau\ll 1$ and $v_Fq\tau\ll 1$, which defines the
diffusive regime, the ladder sum reads
\begin{equation}
\label{diffuson}
L({\bf q},\omega )=\frac{1}{2\pi N_0\tau^2}\frac{1}{Dq^2+|\omega|}.
\end{equation}
The above equation is telling us that the two-particle propagation has
a diffusive form. This means that, whereas the single-particle Green's function
is exponentially decaying over a mean free path, $l =v_F \tau$ ($v_F$ is the 
Fermi velocity), the collective density fluctuations propagate diffusively 
over large distances. It is the long-range diffusive character of the density 
fluctuations which is responsible of the logarithmic behavior of the EEI 
corrections to the conductivity and to the thermodynamic properties in two
dimensions, as we are 
about to see. Technically, the logarithmic corrections arise from the 
integration over the diffusive pole. Each momentum integration, in two 
dimensions, yields a factor $1/D\sim t$, the expansion parameter. We would 
like to emphasize that the emergence of critical massless modes is highly 
non-trivial and shows how the construction of the effective action is in this
case rather unconventional, to say the least.
 
The diffusive form of the two-particle propagator (usually called diffuson) 
makes the small-$q$ region more relevant. As a consequence, the Fourier
transform of the interaction in 
Eq. (\ref{th2}) becomes important only for small momentum transfer, i.e., 
$V(q=0)$. On the other hand, the corresponding Hartree 
diagram contribution, after the average over the impurities, selects the 
interaction at large momentum transfer, i.e., $V(q=2p_F)$. 

The good-metal 
condition, $g\gg 1$, can be written in an equivalent form as 
$E_F \tau /\hbar \gg 1 $, 
which implies that the disorder only affects states within a small shell
$\hbar /\tau$ away from the Fermi surface. Under these circumstances, 
since disorder modifies only states near the Fermi surface, interaction
effects at larger energy can be taken into account via the Fermi-liquid theory,
which goes beyond the first-order perturbation theory in the interaction, 
by replacing $V(q=0)$ and $V(q=2p_F)$ with the corresponding Fermi-liquid 
scattering amplitudes $\Gamma_1$ and $\Gamma_2$, whose lowest order diagrams 
are depicted in Fig. \ref{amplitudes}.
Since, in the absence of spin-flip mechanisms, the total spin entering the
two-particle propagator is a conserved quantity, it is convenient to introduce
the {\sl singlet} and {\sl triplet} scattering amplitudes
$$
\Gamma_s =\Gamma_1-\frac{1}{2}\Gamma_2,~~~~~~~~\Gamma_t =\frac{1}{2}\Gamma_2.
$$
We recall that the scattering amplitudes $\Gamma_s$ and $\Gamma_t$ are related 
to the Landau Fermi-liquid parameters $F_{s,a}^0$ which enter in the
compressibility and spin susceptibility, by
$$
\Gamma_s =\frac{1}{2N_0}\frac{F_s^0}{1+F_s^0},~~~~~~~~
\Gamma_t =-\frac{1}{2N_0}\frac{F_a^0}{1+F_a^0}.
$$

From now on, when necessary, $N_0$ is assumed to include the Landau 
effective-mass correction.  The scattering amplitudes are dynamically 
screened by the diffusive density fluctuations, leading to dressed
scattering amplitudes
$$
\Gamma_{s,t}(q,\omega )=\Gamma_{s,t}
\frac{Dq^2+|\omega|}{Dq^2+(1\mp 2N_0 \Gamma_{s,t})|\omega |}.
$$
However, this dressing does not alter the degree of divergence of a given
diagram.

We are now ready to present the expression for the correction to the 
thermodynamic potential. By means of the previous substitutions, 
this is obtained to all orders in the interaction, but to first order in the 
expansion parameter $t$. It is convenient to use the standard trick \cite{agd} 
of multiplying the interaction 
by a parameter $\lambda$ and integrating over it between $0$ and $1$. We then 
get
\begin{eqnarray*}
\Delta\Omega=&-&T\sum_{q\omega}\int_{0}^{1}{\rm d}\lambda
\left[\frac{N_0\Gamma_s |\omega|}{Dq^2+(1-\lambda 2 N_0\Gamma_s )|\omega|}\right.\nonumber\\
&-&
\left.\sum_M\frac{{N_0 \Gamma}_t |\omega|}{Dq^2+
(1+\lambda 2N_0 \Gamma_t)|\omega|-
{\rm i}M (1+ 2{N_0 \Gamma}_t )\omega_s {\rm sgn} (\omega )}\right].
\nonumber
\end{eqnarray*}
To allow for the calculation of the spin susceptibility, in the above equation 
we have introduced also a magnetic field via the Zeeman 
coupling, $\omega_s =g\mu_B B$ and $M$ labels the triplet states. Notice the 
introduction of the Fermi-liquid renormalization of the Zeeman energy 
\cite{raimondi1990}.  After evaluating 
the integrals one gets
\begin{equation}
\label{th11}
\Delta\Omega =\Delta\Omega_0 +\Delta\Omega_1 +
tN_0T^2\Delta\Omega_2\left( \frac{\omega_s}{T}\right)
\end{equation}
where
\begin{eqnarray}
\Delta\Omega_0&=& -t N_0T^2(2N_0\Gamma_s -6{N_0 \Gamma}_t)
\left( \frac{\pi^2}{6}\ln (T\tau )+A\right), \nonumber\\
\Delta\Omega_1&=& t N_0{N_0 \Gamma}_t(1+2{N_0 \Gamma}_t)
\omega_s^2\ln (\omega_s \tau ),\nonumber\\
\Delta\Omega_2 (x)&=&(1+2{N_0 \Gamma}_t)I(x)-I((1+2{N_0 \Gamma}_t)x)
-4A{N_0 \Gamma}_t,\nonumber\\
I(x)&=&\int_0^{\infty}{\rm d}y b(y)\left[ (y-x)
\ln |y-x| +(y+x)\ln |y+x|\right],
\label{th12}
\end{eqnarray}
where $b(x)$ is the Bose function and $A\approx -0.24$. Eqs.
(\ref{th11},\ref{th12}) are organized to show the zero- and the 
large-magnetic-field behaviors, which are connected by the crossover function 
$\Delta\Omega_2 (x)$. Notice that at small $x$, one has
$\Delta\Omega_2 (x)\approx  -{N_0 \Gamma}_t(1+2{N_0 \Gamma}_t)x^2\ln x$.
As a result, the corrections to the specific heat and to the spin 
susceptibility read
\begin{eqnarray}
\delta C_V&=&C_{V,0}t(N_0\Gamma_s -3{N_0 \Gamma}_t)\ln (T\tau ),\nonumber\\
\delta \chi&=&-\chi_0 4t{N_0 \Gamma}_t(1+2{N_0 \Gamma}_t)
\ln (T\tau ),\label{th13}
\end{eqnarray}
where $C_{V,0}=2\pi^2 N_0T/3$ and $\chi_0 =N_0 (g\mu_B )^2/2$ are the
non-interacting values. We note that there is no dependence on the chemical 
potential implying that there are no singular corrections to the static 
compressibility $\partial n/\partial \mu$, i.e., 
$\delta (\partial n/\partial \mu )=0$. As a result, the compressibility has 
the Landau value
\begin{equation}
\frac{\partial n}{\partial\mu}=N_0\left(1-2N_0\Gamma_s\right)\equiv
N_0 Z_s^0.
\label{landaucompr}
\end{equation}

Eq.(\ref{th13}) contains the leading divergent perturbative terms
to  thermodynamic  properties of a disordered Fermi liquid.
Before proceeding to the next sections, where we discuss the renormalized 
perturbation theory  and how to obtain the equations for the RG flow including transport
properties, we remark 
that in various physical systems the above corrections can actually be 
observed already in the metallic phase, where they are still within the reach 
of perturbation theory \cite{ando1987}. 

\subsection{Ward identities and renormalized Fermi liquid}
In order to get the RG equations, one has to absorb the logarithmic
corrections in terms of the renormalization of a set of running variables.
This is usually done in terms of the relevant parameters of the Hamiltonian.
The difficulty of the problem at hand is precisely the fact that the effective 
Hamiltonian is not simply related to the microscopic one we started from. We 
have seen in the previous section that the logarithmic correction originates 
from an integration over the collective diffusive density modes. The 
propagator of the diffusive mode, which we have called ladder, is the result 
of a re-summation of impurity scattering to all orders. If we are going to 
obtain a renormalized theory, we must understand how the logarithmic 
correction we have obtained in the previous section will manifest in the 
effective field theory for the diffusive modes. This task can actually be 
carried out systematically from the outset by deriving a matrix-field 
non-linear $\sigma$-model \cite{finkelstein1983}. This approach, however, 
makes the connection with the physical quantities less direct. By the use of 
the Ward identities, all the renormalizations required by the non-linear 
$\sigma$-model are expressed through the Landau parameters, thus showing that 
the disordered interacting electron system can be effectively described by a 
scale-dependent Landau Fermi-liquid theory 
\cite{altshuler1983b,castellani1984,castellani1986,castellani1986b,castellani1987}.
For this 
reason, we adopt here the approach of perturbation theory combined with the 
Ward identities.

We begin by recalling the expression of the density and current in linear
response regime
$$
j_{\mu}(\omega ,{\bf q})=K_{\mu\nu}(\omega ,{\bf q})A_{\nu}(\omega ,{\bf q}),
$$
where $j_{\mu}=(\rho , {\bf j})$ and $A_{\mu}=(\phi , {\bf A})$.
$K_{\mu\nu}$ is the standard response function for density and current. In 
particular the conductivity and the compressibility are given by
\begin{eqnarray*}
\sigma&=&-\lim_{\omega\rightarrow 0}\frac{{\rm Im}K}{\omega},\nonumber\\
\frac{\partial n}{\partial\mu}&=&-\lim_{{\bf q}\rightarrow 0}
K_{00}(0, {\bf q}),\nonumber
\end{eqnarray*}
where $K=(1/d)\sum_{1,d}K_{ii}$, assuming spatial isotropy. In a similar way, 
one defines the spin-spin  $\chi(\omega , {\bf q})$ and energy-energy 
$\chi_E(\omega , {\bf q})$ response functions, whose static limit (i.e., 
$\omega\rightarrow 0$ first, ${\bf q}\rightarrow 0$ after) are the spin 
susceptibility and the specific heat times the temperature.

The global conservation law and gauge invariance give rise to the following
Ward identities 
\begin{eqnarray}
q_{\mu}K_{\mu\nu}&=&0,\nonumber\\
K_{\mu\nu}q_{\nu}&=&0,
\label{wi3}
\end{eqnarray}
that must be obeyed to all orders in perturbation theory. Notice that the 
first equation is nothing but a different way of writing Eq. (\ref{usualwi}). 
For instance, a disordered non-interacting Fermi system has a density-density 
response function given by
$$
K_{00}=-\frac{\partial n}{\partial\mu}\frac{Dq^2}{Dq^2-{\rm i}\omega}
=-\frac{\partial n}{\partial\mu}-\frac{{\rm i}\omega N_0}{Dq^2-{\rm i}\omega},
$$
The first expression is nothing but a statement about the 
diffusive nature of the density fluctuations. The second expression, which 
separates the so-called dynamic part of the response function, is the result 
of a direct diagrammatic calculation.
To understand this point, recall that the evaluation of the bubble-like
diagram involves an integral over the energy of a product of two
Green's functions. The dynamic part corresponds to the energy range 
where the two Green's functions have poles on opposite sides of the real axis,
in the complex plane, and a ladder resummation can be performed.
$N_0$ appears since we are 
dealing with a non-interacting system. From Eqs. (\ref{wi3}), it follows that
\begin{equation}
\label{wi5}
K_{00}(\omega , 0)=0,
\end{equation}
which immediately gives that the compressibility $\partial n/\partial\mu$ 
coincides with the single-particle density of states, as it is expected for 
the non-interacting system. The single-particle density of states, 
furthermore, may be generally related 
to the dynamic part of the response function $K^{+-}_{00}$, i.e.,
\begin{equation}
\label{dynamicpartofk00}
K^{+-}_{00}=-{\rm i}\int_{-\omega}^0\frac{{\rm d}\varepsilon}{2\pi}
\sum_{\bf p}G({\bf p},\varepsilon )G({\bf p}+{\bf q},\varepsilon +\omega )
\Lambda_0^{+-} ({\bf p},{\bf q},\varepsilon , \omega ),
\end{equation}
where $\Lambda_0^{+-}$ is the density vertex  in the energy range
$\epsilon (\epsilon +\omega )<0$, which in the non-interacting
case gives the total dynamic part of the response function.
The reason for defining the 
vertex $\Lambda_0^{+-}$ in this energy range is due to the possibility of inserting the
ladder resummation.
 For positive external frequency $\omega$,
 the energy restriction determines  the integration range over the energy $\epsilon$
 in Eq. (\ref{dynamicpartofk00}). 
By using the general 
Ward identity Eq. (\ref{usualwi}), in the dynamic limit, one gets
the single-particle density of states
\begin{equation}
\label{wi6}
N\equiv -\frac{1}{\pi}\sum_{\bf p}{\rm Im}G({\bf p},\varepsilon ={\rm i}0^+)
=\lim_{\omega\rightarrow 0}
\lim_{{\bf q}\rightarrow 0}K^{+-}_{00}(\omega ,{\bf q}).
\end{equation}
In the simple non-interacting case, of course, $N=N_0$. 

We stress that 
both Eqs. (\ref{wi5},\ref{wi6}) are always valid and we use them in the 
following when dealing with interacting electrons.
Here, they will allow us to control the logarithmic corrections which arise
from the interplay 
between the diffusive motion of the electrons and their mutual interaction. 
 We refer to renormalized quantities 
with a bar. The diagrammatic skeleton structure of the response function is shown
in Fig. \ref{responsefunction}. 
$\Lambda_s$ is, in the interacting case, the vertex which, 
 when multiplied by $K_{00}^{+-}$ 
gives the total dynamic part of $K_{00}$, which includes also terms ending with
two advanced ($++$) or retarded ($--$) Green's functions\cite{castellani1986b}. 
A direct perturbative evaluation of the density response function 
gives
\begin{equation}
\label{wi7}
{\bar K}_{00}^{+-} = 
-\frac{{\rm i}\omega N_0\zeta^2\Lambda_s}{{\bar D}q^2-
{\rm i}\omega Z_s},~~~~~~~~
{\bar K}_{00} = -\overline{\frac{\partial n}{\partial\mu}}
+{\bar K}_{00}^{+-}\Lambda_s
\end{equation}
where 
\begin{equation}
\label{zs}
Z_s \equiv Z-2N_0\zeta^2{\bar \Gamma}_s
\end{equation}
is the renormalized form of
$Z_s^0$, defined in Eq. (\ref{landaucompr}), and the singular corrections have 
been absorbed in the renormalized quantities 
${\bar D}$, $Z$,  $\zeta $, ${\bar \Gamma}_s$ and $\Lambda_s$. The 
first three are obtained by considering the corrections to the ladder in the 
presence of interaction. ${\bar \Gamma}_s$ is the renormalized scattering 
amplitude in the singlet channel. 

Here, the parameter $\zeta$  plays the role of the 
{\sl wave-function} renormalization of the effective field theory with the 
ladder as a propagator, while $Z$ renormalizes the frequency, and it will be
identified with the specific heat renormalization parameter.
 By the use of Ward identities we now show that $\zeta$  also 
coincides with the renormalization of the single-particle density of states 
introduced by the disorder, in the presence of the interaction, i.e.,
\begin{equation}
\label{wi8}
N=\zeta N_0.
\end{equation}
Indeed, when Eqs. (\ref{wi5},\ref{wi6}) are used in Eq. (\ref{wi7}), we
obtain
\begin{equation}
\frac{N}{N_0}=\frac{\zeta^2\Lambda_s}{Z_s},~~~~~~~~
\overline\frac{\partial n}{\partial \mu}=N_0\frac{\zeta^2\Lambda_s^2}{Z_s}.
\label{wi9}
\end{equation}
Since the compressibility has no singular corrections, $Z_s$ coincides with 
$Z_s^0$ and $\zeta\Lambda_s=Z_s^0$, whence Eq. (\ref{wi8}) follows. Thus, the 
single-particle density of states becomes scale-dependent, in contrast to the 
non-interacting case.

In a similar way, considering $Z_t=Z-2N_0\zeta^2\bar\Gamma_t$, one
obtains for the specific heat and for the spin susceptibility  
the relations
\begin{eqnarray}
{\bar C}_V&=&Z C_V,\nonumber\\
{\bar \chi}&=&Z_t N_0.
\label{wi12}
\end{eqnarray}
The above equations relate the renormalization parameters $Z$, and
$Z_t$ that appear in the 
perturbative correction to the physical quantities $C_V$ and $\chi$. Contrary to $Z_s$, $Z$ 
and $Z_t$ are affected by logarithmic singularities, which can be derived
through the evaluation of the thermodynamic potential. Alternatively, one can 
compute the corrections to $Z$, and $Z_t$ directly (although with a lot more 
effort) in perturbation theory and find a perfect agreement between the 
expressions obtained via the thermodynamic potential and Eqs.
(\ref{wi12}). 

Before going to the next section, to deal with the RG equations,  we
would like to recall the complete expression for the electrical
conductivity. To do so, we notice that 
in the presence of Coulomb long-range forces, to avoid double counting,
one has
to subtract the statically screened long-range Coulomb $\Gamma_0$ from
the full singlet scattering amplitude entering the ladder resummation for
the density-density response function.
Hence, 
${\bar{\Gamma_s}} \rightarrow {\bar{\Gamma_s}} -\Gamma_0$,  Eq.(\ref{zs})
for $Z_s$ is modified accordingly, and $\Gamma_0$ reads
\begin{eqnarray} 
\Gamma_0 (q,  \omega =0)&=&\frac{V_C (q)\Lambda_s^2}{1+V_C(q)2{\bar{\partial n /\partial \mu}}}
 \rightarrow_{q\rightarrow 0}
\frac{\Lambda_s^2}{2{\bar{\partial n /\partial \mu}}}.
\end{eqnarray}
As a consequence, by using Eq.(\ref{wi9}), 
we derive the constraint
\begin{equation}
\label{wi13}
Z=2N_0\zeta^2{\bar \Gamma}_s.
\end{equation}
Then the EEI correction to the 
conductivity by including also the effect of the triplet channel and of the 
dynamical re-summation \cite{finkelstein1983,castellani1984} becomes
\begin{equation} 
\label{th14}
\delta t =-t^2\left[ 1+3\left(1-\frac{(1+2{\gamma}_t)}
{2{\gamma}_t}\ln (1+2{\gamma}_t)\right)\right]\ln (T\tau ),
\end{equation}
where $ N_0\zeta^2{\bar \Gamma_t} /Z=\gamma_t$.
The first term in the square brakets is the contribution due to the
singlet channel in the presence of long-range Coulomb interaction
[see Eq. (\ref{interactioncorrection})]. 
Notice that  the corresponding scattering amplitude
does not appear explicitly, due to the constraint (\ref{wi13}).
 An analogous contribution would come from the WL
correction, given by Eq. (\ref{weaklocalization}). This last term, 
together with the contribution of the interaction in the
particle-particle channel, is
here suppressed to maintain the discussion simple while keeping all the
general features of the theory.

Finally, we would like 
to  remark about the various ``density of states'' we have 
encountered. We have the thermodynamic density of states or compressibility 
$\partial n/\partial\mu$, the coefficient of the linear-in-$T$ term of the 
specific heat, $\gamma$, and the single-particle density of states obtained 
from the Green's function, $N$. In the non-interacting case all these quantities,
together with the spin susceptibility, coincide with $N_0$, but in the 
presence of interaction and disorder they are renormalized by $Z_s$, $Z$, 
$\zeta$, and $Z_t$, respectively. The Ward identities are a very powerful tool 
in controlling the different role played by these quantities in the 
renormalized theory.

\subsection{One-loop RG equations for the renormalized Fermi liquid}
\begin{figure}[h]
\resizebox{.5\textwidth}{!}
{\includegraphics{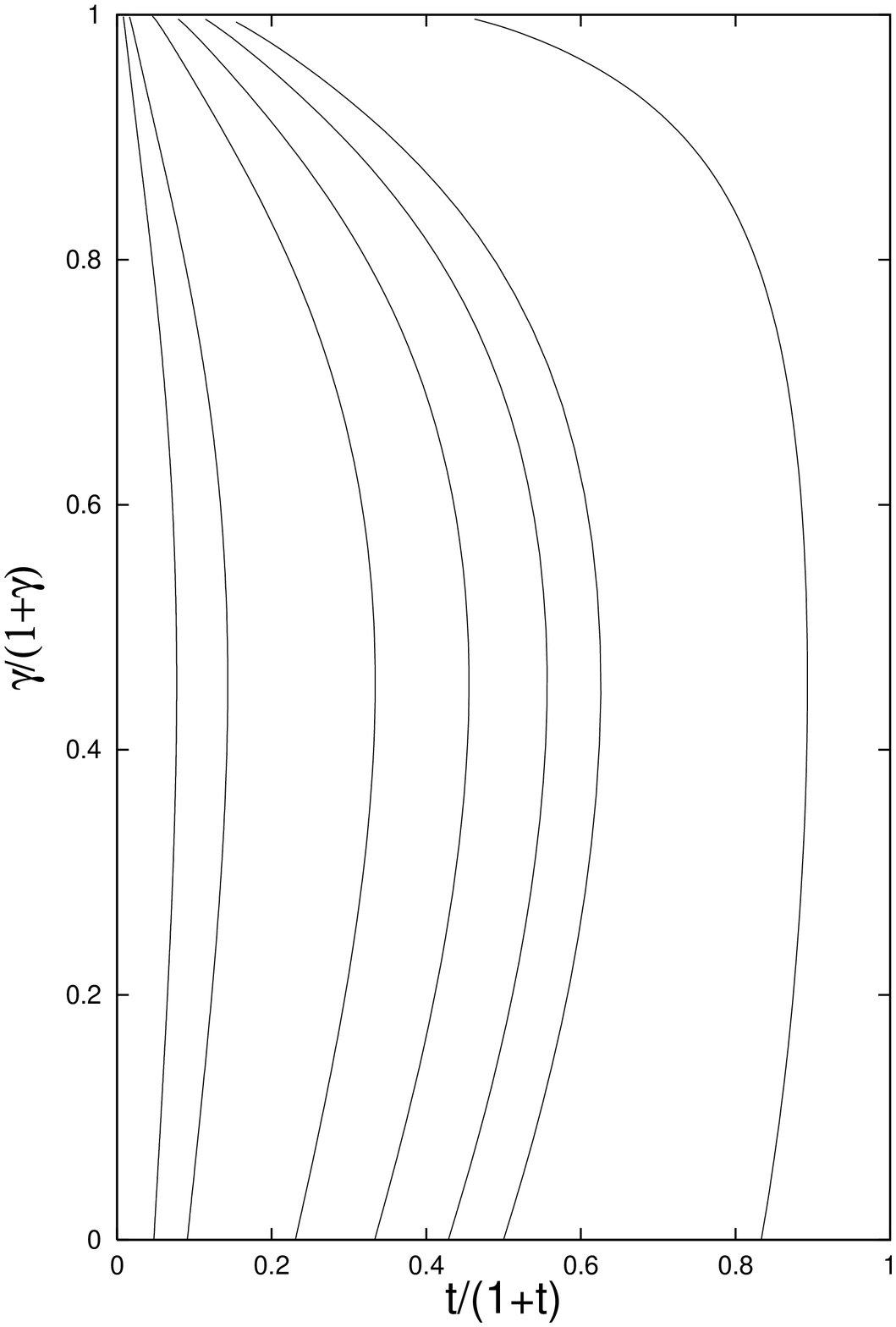}}
\caption{C. Di Castro, {\sl et al.} - Renormalization group...}
\label{rgflow}
\end{figure}

In the previous section we have seen that all the divergences which appear in 
perturbation theory may be absorbed into the renormalization of the physical 
parameters characterizing a Fermi liquid. Therefore, we can use the previous
results to obtain  here the RG flow equations for $Z=C_V/C_{V,0}$ and 
$Z_t/Z=\chi/Z\chi_0$. From Eqs. (\ref{th13},\ref{th14}), introducing 
the scaling variable $s=-\ln (T\tau)$, one gets
\cite{finkelstein1983,castellani1984,castellani1984b,finkelstein1984,castellani1986b}
\begin{equation}
\label{th19}
\left\{
\matrix{
\displaystyle{\frac{{\rm d}Z}{{\rm d}s}}&=&
\displaystyle{-\frac{tZ}{2}(1-6{\gamma}_t),
~~~~~~~~~~~~~~~~~~~~~~~~~~~~~~~~~~~~~~~~~~~}\cr
\displaystyle{\frac{{\rm d}{\gamma}_t}{{\rm d}s}}&=& 
\displaystyle{\frac{t}{4}(1+2{\gamma}_t )^2,
~~~~~~~~~~~~~~~~~~~~~~~~~~~~~~~~~~~~~~~~~~~~}\cr
\displaystyle{\frac{{\rm d}t}{{\rm d}s}}&=&
\displaystyle{-\epsilon \frac{t}{2}+
t^2\left[ 1+3\left(1-\frac{(1+2{\gamma}_t)}
{2{\gamma}_t}\ln (1+2{\gamma}_t)\right)\right],}
}\right.
\end{equation}
where the last equation is derived through Eq. (\ref{th14}),
considering the bare dimension $\epsilon=d-2$ of the coupling $t$ in
units of inverse length. 
We briefly review the main consequences of the above equations. Let us begin 
the discussion in two dimensions, i.e., $\epsilon =0$
(see Fig.  \ref{rgflow}). Under the RG, 
${\gamma}_t$ grows always and diverges at a finite scale.
 In the equation for $t$ the singlet 
contribution makes it larger (localizing character), whereas the
triplet contribution (in the round brackets) does 
the opposite (antilocalizing). If the initial value of ${\gamma}_t$ is 
not too large, $t$ initially grows until the growth of ${\gamma}_t$ 
makes the triplet contribution the dominating one. As a result, $t$ has a 
maximum as a function of $s$. However, due to the strong-coupling runway of 
${\gamma}_t$, one cannot seriously trust the above equations 
quantitatively. Nevertheless, the physical indication of some type of 
ferromagnetic instability is rather clear due to the diverging spin
susceptibility associated with $\gamma_t$. Furthermore,  the
dominating antilocalizing effect of the triplet while $t$ remains finite,
strongly supports the possibility of a 
metallic phase at low temperature, in contrast with the non-interacting theory
based on WL only
\cite{castellani1984b,finkelstein1984,castellani1986b,castellani_98,nussi}.
Indeed, this metallic phase in $d=2$ has been recently observed
(See Refs. in \cite{abrahams2000}). 
In addition, due to the divergence of $\gamma_t$ also $Z$ goes to the strong coupling regime,
leading to an enhancement of the specific heat, which is however hardly
observable in two dimensions.

\begin{figure}[h]
\resizebox{.5\textwidth}{!}
{\includegraphics{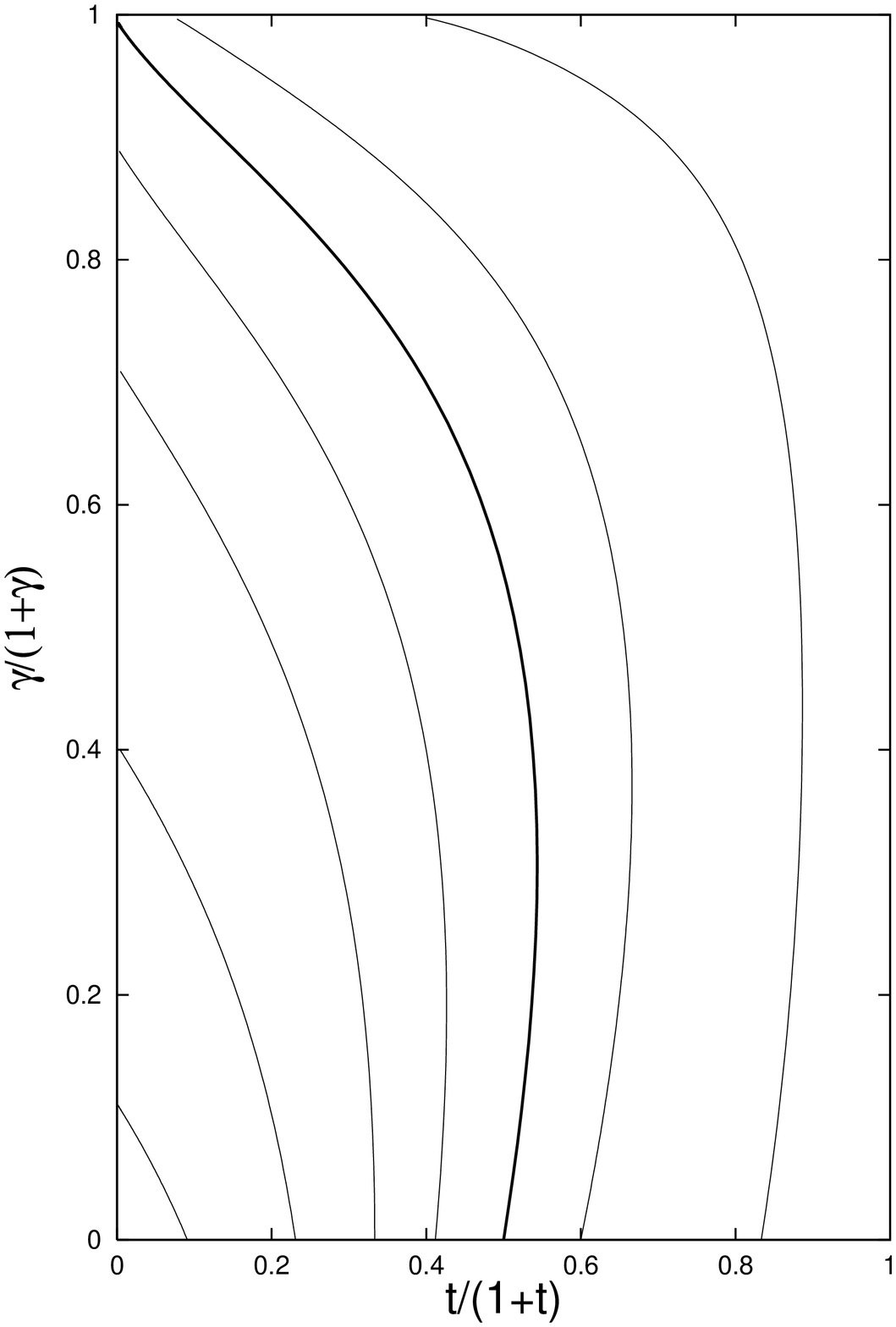}}
\caption{C. Di Castro, {\sl et al.} - Renormalization group...}
\label{rgflow3}
\end{figure}

In three dimensions 
(see Fig. \ref{rgflow3}) one has a richer scenario depending on the 
initial values of the running couplings. The main difference is that there 
exists a critical line in the $t-\gamma_t$ plane asymptotically given by 
$t\gamma_t =\epsilon/2$, where under the RG flow $t\rightarrow 0$, 
$\gamma_t\rightarrow \infty$ with their product being constant. On the 
weak-disorder side of this line, the system scales to a  conductor
with vanishing $t$, whereas on 
the strong-disorder side the system behaves qualitatively as in the two-dimensional 
case discussed above, leaving again the possibility open for a ferromagnetic 
instability at a finite scale. In this latter case, however, the strong-coupling runaway flow 
requires to go beyond the one-loop approximation we have presented here
leaving  the problem of the proper treatment of the metal-insulator
transition still open. An 
approximate treatment of the two-loop correction is possible, but its 
discussion is well outside the scope of this paper. We refer the reader to 
Ref. \cite{belitz1994}.
However, we remark   that, whenever there is a 
reduction of the effect of the triplet channel, due to a
magnetic coupling altering the symmetry, 
as in the presence of a magnetic field with Zeeman spin-splitting
\cite{castellani1984,finkelstein1984b}, 
spin-flip scattering\cite{altshuler1983b,castellani1984,finkelstein1984b} or spin-orbit 
scattering \cite{altshuler1983b,castellani1984c} one obtains a {\it bona-fide} MIT with different 
universality classes with respect to the Anderson localization. 
For a comparison with the experiments see, e.g., Ref. 
\cite{dicastro1987}.

We finally emphasize that, as in the interacting boson case, the perturbative
RG at one loop confirms the conclusions drawn by the implementation of the
Ward identities. However, in the present case, 
since we are dealing with critical behavior,
the singularities are resummed in power laws, 
rather than to be cancelled  as required in a stable liquid
phase. Mathematically this manifests in the lack of additional symmetries
which prevents us from closing the equations of motion and
 solving  the problem exactly.  We have achieved, however, via the Ward identities, that
the remaining singularities describe
the critical behavior of the specific heat, spin susceptibility, and 
of the coupling $t$ related to the conductance.

\vskip 1truecm
\par\noindent{\bf Acknowledgments.} S. C. and C. D. C. acknowledge financial
support from the Italian MIUR, Cofin 2001, prot. 20010203848, and from INFM,
PA-G0-4. R. R. acknowledges partial financial support from E.U. 
by Grant RTN 1-1999-00406, and from the Italian MIUR, Cofin 2002.

\newpage

\centerline{\bf Table captions}

Tab. \ref{table1}. In the table we give a summary of the conservation laws 
exploited in order to keep under control the singularities arising in 
perturbation theory, for each of the physical systems discussed in the text. 
In particular, we indicate the conservation laws (second column), the 
practical consequence of their implementation (third column), the behavior
of the system (fourth column), and the equation number where the 
corresponding Ward identity can be found in the article (fifth column).

\newpage
\begin{table}
\begin{tabular}[]{|p{1.75cm}|p{2cm}|p{2cm}|p{1.5cm}|p{2cm}|} 
\hline
Physical System & Conservation Law& Consequence & Behavior &Ward Identity\\
\hline 
\hline
Interacting Fermions & Global and separate Left and Right number density&
Cancellation of singularities in 
correlation functions & Unusual non-critical& Eqs. (\ref{usualwi},\ref{newwi}) 
\\ \hline 
Interacting Bosons & Global number density and $U(1)$-gauge invariance&
Identification and reduction of running variables
&Unusual non-critical &Eq. (\ref{bosesymmetry})\\ \hline
Disordered Interacting Electrons & Global number density and $U(1)$-gauge 
invariance&
Identification and reduction of running variables &Critical &
Eqs. (\ref{usualwi},\ref{wi3})\\ \hline
\end{tabular} 
\caption{C. Di Castro, {\sl et al.} - Renormalization group...}
\label{table1} 
\end{table}

~\newpage

\centerline{\bf Figure captions}

\begin{figure}[h]
\resizebox{.8\textwidth}{!}
{\includegraphics{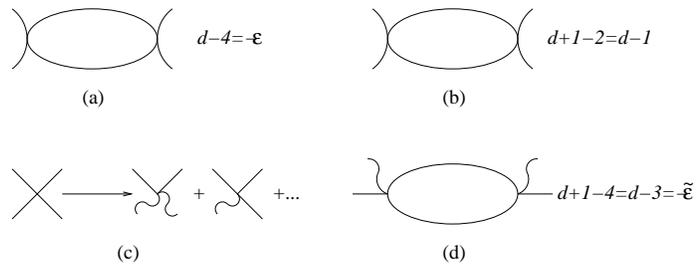}}
\caption{C. Di Castro, {\sl et al.} - Renormalization group...}
\label{diagr}
\end{figure}

Fig. \ref{diagr}. (a): Lowest-order correction to the coupling constant of the
$\varphi^4$ theory. Each line in the loop carries a factor
$[t+{\bf q}^2]^{-1}$, and an integral over $d^d{\bf q}$ is implied, giving a
contribution which diverges as $|t|^{-\epsilon/2}$ for $t\to 0$, when
$\epsilon>0$, i.e., $d<4$. The divergence is logarithmic for $\epsilon=0$
($d=4$).
(b): Lowest-order correction to the coupling constant of an interacting Fermi
system. Each line in the loop carries a factor
$[\varepsilon-v_F(|{\bf p}|-p_F)]^{-1}$, where $v_F$ is the Fermi velocity
and $p_F$ the Fermi momentum, and an integral over $d\varepsilon d^d{\bf p}$
is implied, giving a logarithmic divergence in the external momentum or
frequency, when $d=1$.
(c): The four-point vertex of an interacting Bose systems generates also two-
and three-leg vertices below the condensation temperature. The dashed line
represents a condensed boson.
(d): Lowest-order correction to the single-particle propagator of an
interacting Bose system, associated with three-leg vertices below the
condensation temperature. Each line in the loop carries a factor
$[\omega^2-c^2{\bf q}^2]^{-1}$, and an integral over $d\omega d^d{\bf q}$ is
implied, giving a divergence as $s^{-\tilde\epsilon}$, where $s$ is a finite
external momentum or frequency, for $\tilde\epsilon>0$, i.e., $d<3$.
The divergence is logarithmic for $\tilde\epsilon=0$ ($d=3$).

\begin{figure}[h]
\resizebox{.8\textwidth}{!}
{\includegraphics{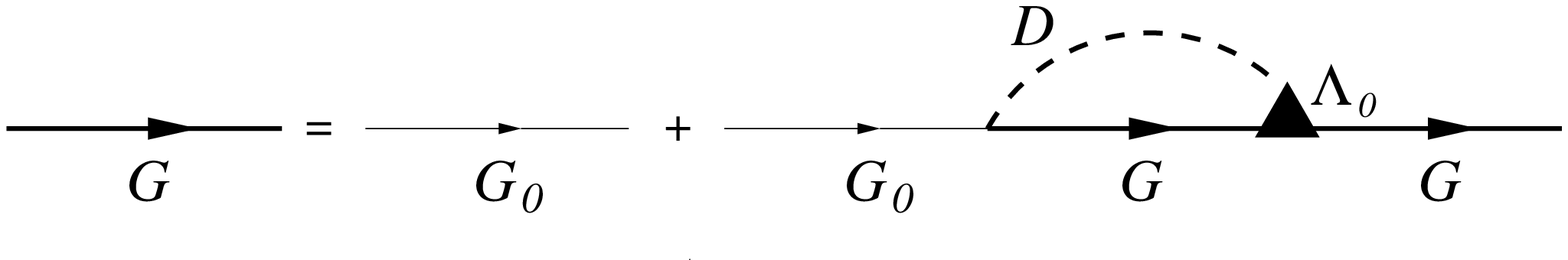}}
\caption{C. Di Castro, {\sl et al.} - Renormalization group...}
\label{dysonfig}
\end{figure}

Fig. \ref{dysonfig}. Dyson equation for the electron Green's function $G$ (thick 
solid line). The thin solid line represents the bare Green's function $G_0$,
the tick-dashed line is the effective interaction $D$ (dressed by the RPA series),
and the black triangle represents the irreducible scalar vertex $\Lambda_0$.

\begin{figure}[h]
\resizebox{.8\textwidth}{!}
{\includegraphics{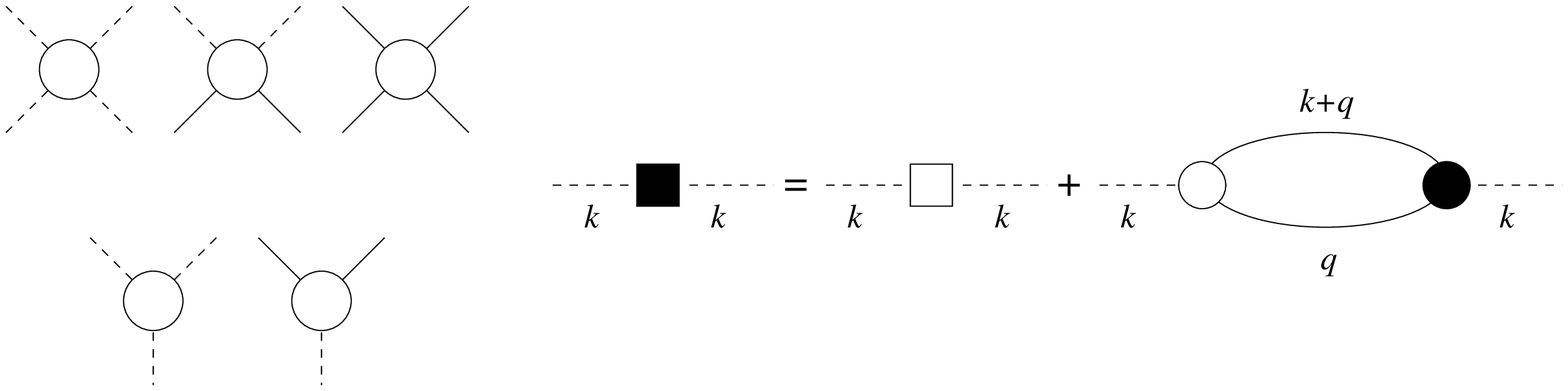}}
\caption{C. Di Castro, {\sl et al.} - Renormalization group...}
\label{figbos}
\end{figure}

Fig. \ref{figbos}. Left: Various four- and three-leg vertices arising in the 
broken-symmetry phase. Dashed and solid lines represent longitudinal ($L$) and
transverse ($T$) modes, respectively.
Right: Equation for the most singular part of the two-point vertex $v_{LL}$ 
(filled square). The empty square represents the bare vertex $v_{LL}^0$, 
whereas the empty and black circles represent the three-leg vertices 
$v_{LTT}^0$ (bare) and $v_{LTT}$ (dressed), respectively. The external dashed 
lines, which represent longitudinal fluctuations, are amputated, and are only 
drawn to indicate the corresponding ingoing and outgoing momenta. The solid 
line represent the propagator of transverse fluctuations, $G_{TT}$. Notice 
that the perturbative one-loop RG equation for $v_{LL}$
is obtained by taking the bare vertices only.

\begin{figure}[h]
{\includegraphics{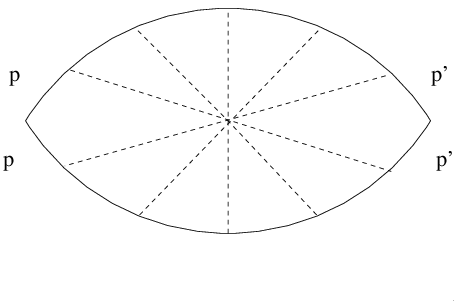}}\hskip 1cm
{\includegraphics{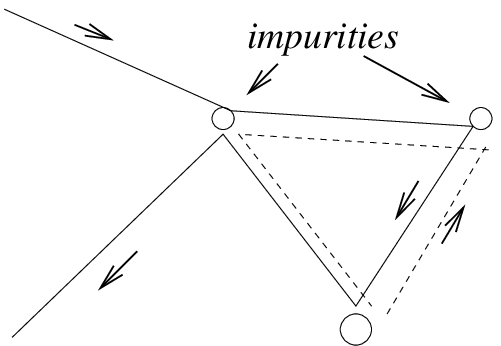}}
\caption{C. Di Castro, {\sl et al.} - Renormalization group...}
\label{weaklocalization_fig}
\end{figure}

\begin{figure}[h]
\resizebox{.8\textwidth}{!}
{\includegraphics{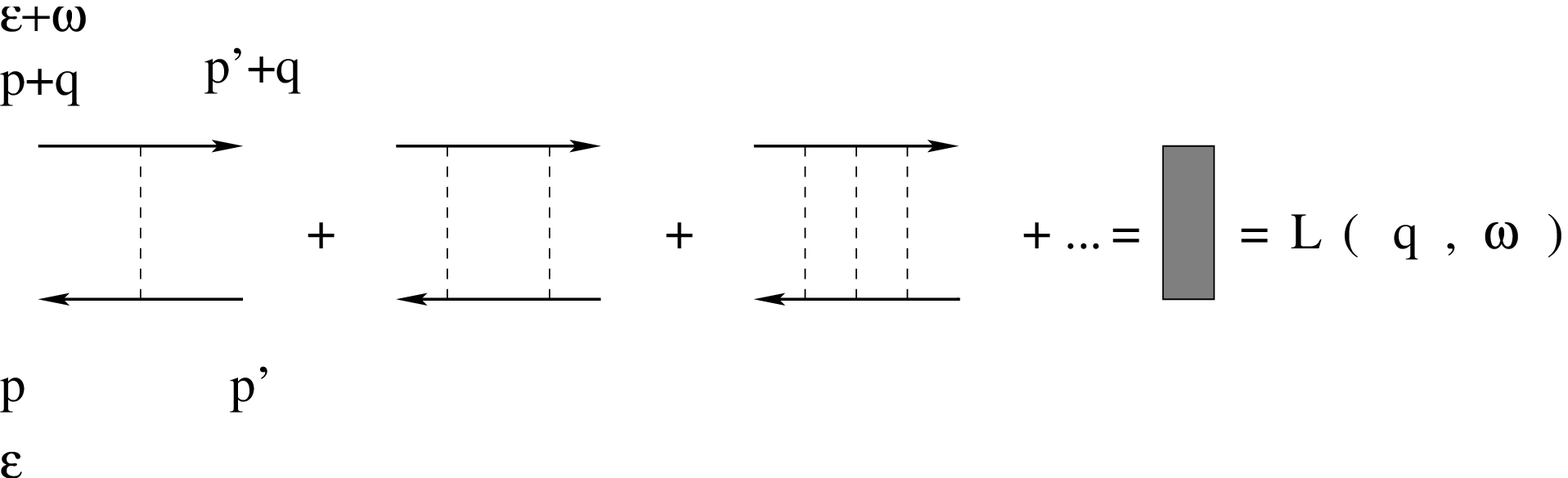}}
\caption{C. Di Castro, {\sl et al.} - Renormalization group...}
\label{ladder}
\end{figure}

Fig. \ref{weaklocalization_fig}. Left: Diagram giving the WL correction to the 
conductivity, Eq. (\ref{weaklocalization}). The dashed line represents the 
average of two impurity 
insertions, introduced later on in the text, Eq. (\ref{th3}). 
In general, it is necessary to sum up diagrams with an infinite number
of impurity lines. The leading term gives rise to the ladder diagram explained
in the text [cf. Eq. (\ref{diffuson}), see also Fig. \ref{ladder}]. 
In the case of electrical conductivity, one can show
that this leading term reproduces the semiclassical result of the Drude-Boltzmann
theory. The next correction is obtained by considering the  series of 
maximally crossed ladder diagrams shown here. In Ref. \cite{gorkov1979},
it has been shown that the crossed ladder may be evaluated in terms of the
{\sl direct} ladder by exploting the time reversal symmetry. 
Notice that, because of the crossing of all the impurity lines,
the impurities visited by the top Green's function line are met in reversed
order by the bottom one. This is the diagrammatic representation
of the interference of time-reversed trajectories.
Right: 
The WL correction  due to the interference by pairs of trajectories, one 
the time reversed of the other, is shown in a more pictorial way. 
In the figure, the two trajectories  
(solid and dashed lines) differ in the way one goes around the closed loop.

\begin{figure}[h]
{\includegraphics{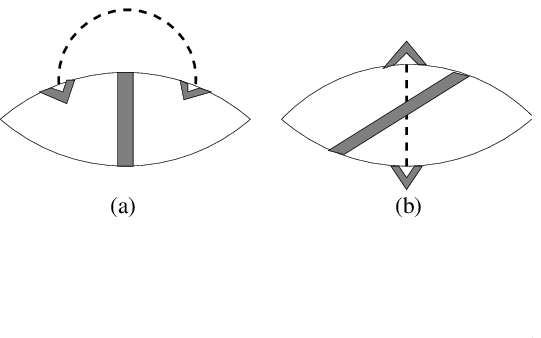}}\hskip 1cm
{\includegraphics{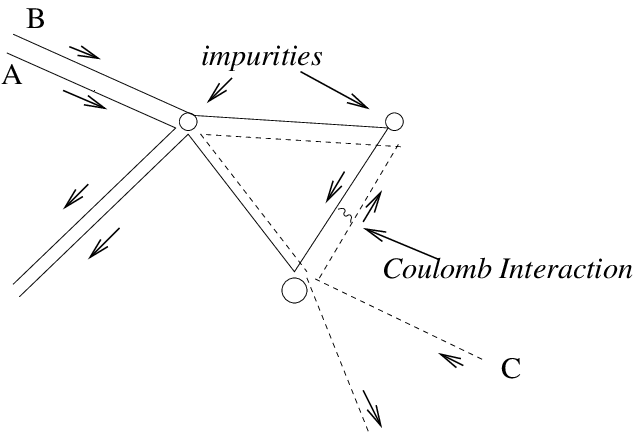}}
\caption{C. Di Castro, {\sl et al.} - Renormalization group...}
\label{effectiveconductivitydiagrams}
\end{figure}

Fig. \ref{effectiveconductivitydiagrams}. Left: 
After non-trivial 
cancellations, as explained in Ref. \cite{altshuler1980}, the diagrams 
(a) and (b) are those responsible for the EEI correction to the conductivity,
Eq. (\ref{interactioncorrection}).
The diagrams shown here contribute to the singlet channel. Another set 
of diagrams,
not shown here, may be obtained by considering the Hartree-like contributions.
This latter set of diagrams gives rise to the correction to the conductivity
in the triplet channel.
In the figure, the tick dashed line represents the electron-electron 
interaction and the shaded parts the insertions of the infinite series
of ladder diagrams[ see also Fig. \ref{ladder}].
Right: A 
physical picture of the processes taken into account by the diagrams on the 
left. Two trajectories (A and B) that differ by a closed loop do not, in 
general, interfere. The electron-electron interaction, however, may lead to 
interference, since the extra phase gained in the closed loop may be canceled 
by another electron (C) going around the same closed trajectory.

\begin{figure}[h]
\resizebox{.5\textwidth}{!}
{\includegraphics{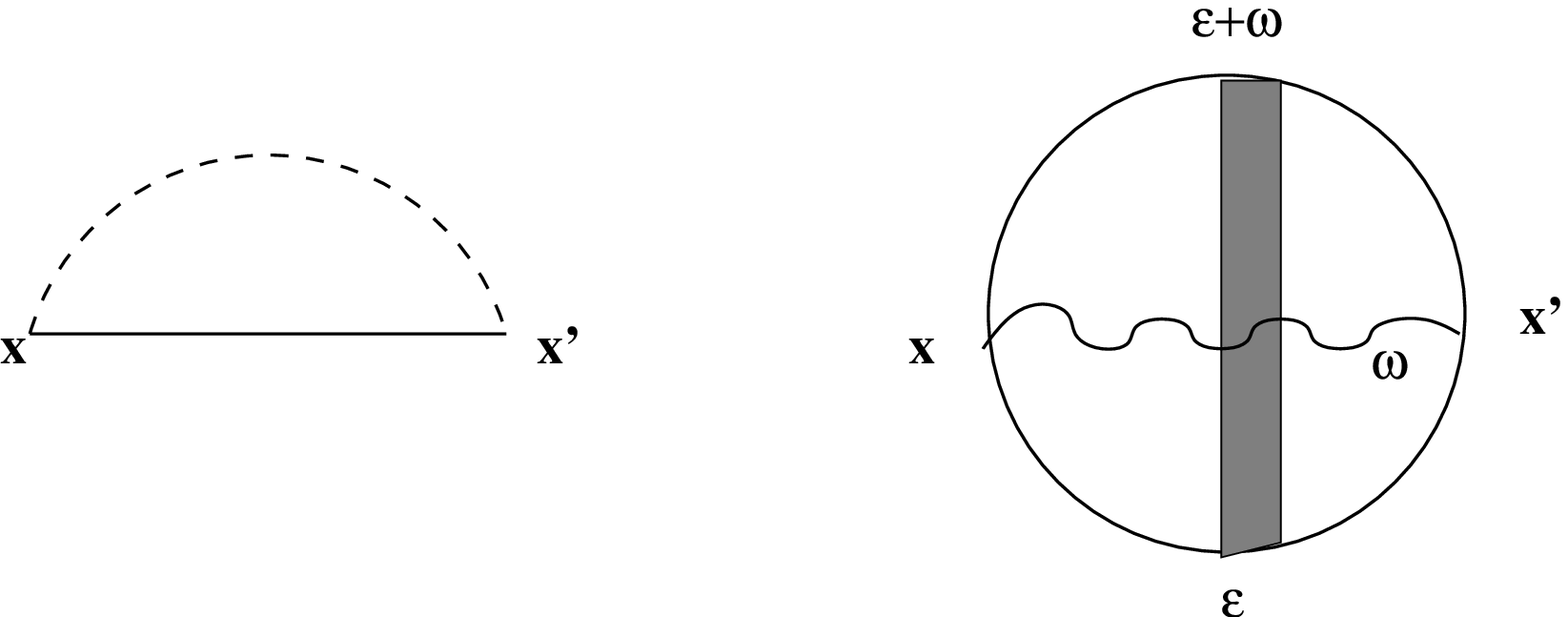}}
\caption{C. Di Castro, {\sl et al.} - Renormalization group...}
\label{thermo}
\end{figure}

Fig. \ref{thermo}. Left: Self-energy in the Born approximation. The dashed 
line represents the average of two impurity insertions, Eq. (\ref{th3}).
When the internal 
Green's function (solid line) is replaced with the dressed Green's function one 
obtains the self-consistent Born approximation. Right:
 First-order correction to the thermodynamic potential (exchange term).
Upon impurity averaging, the ladder re-summation appears,
which corresponds to repeated independent scattering events (see Fig.
\ref{ladder})

Fig. \ref{ladder}. The diagrammatic series of repeated impurity scattering. 
The dashed
line represents the average over the impurity strength distribution.
Physically, both the top (electron) and bottom (hole) Green's function lines
visit the same impurity site. Notice that, since the impurity scattering is
elastic, the energy is not changed along an electron line.
The infinite series of ladder diagrams is indicated with a shaded rectangle.

\begin{figure}[h]
\resizebox{.8\textwidth}{!}
{\includegraphics{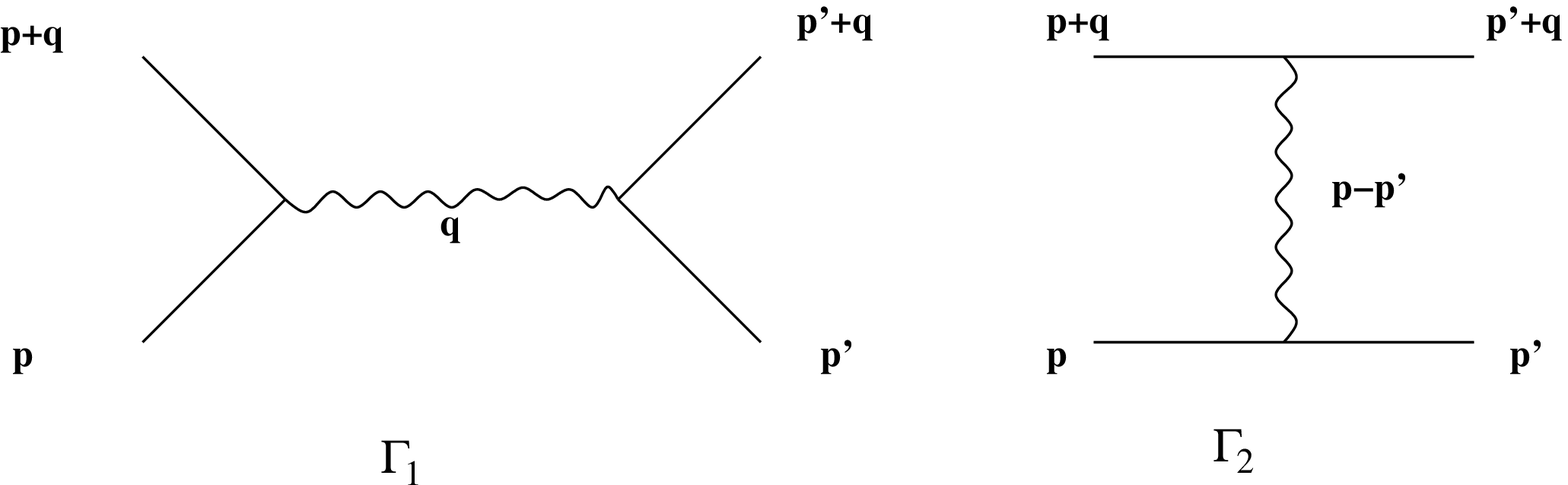}}
\caption{C. Di Castro, {\sl et al.} - Renormalization group...}
\label{amplitudes}
\end{figure}

Fig. \ref{amplitudes}. Lowest order diagrams for small ($\Gamma_1$)
and large  ($\Gamma_2$ ) momentum transfer (scattering angle).

\begin{figure}[h]
\resizebox{.8\textwidth}{!}
{\includegraphics{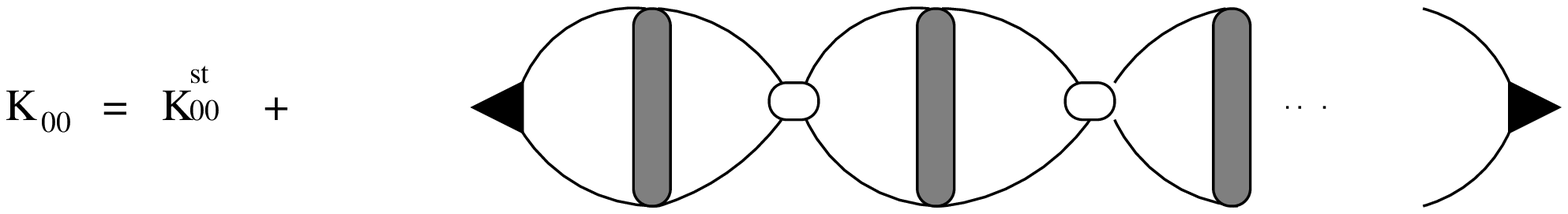}}
\caption{C. Di Castro, {\sl et al.} - Renormalization group...}
\label{responsefunction}
\end{figure}

Fig. \ref{responsefunction}. Skeleton diagram of  the response
function. The black triangles represent the  vertex $\Lambda_s$, the
open squares the scattering amplitudes $\Gamma_{s,t}$, and the shaded rectangle
the ladder of impurity lines, Eq. (\ref{diffuson}). $K_{00}^{st}$ is the 
static limit
of the response function, i.e., the compressibility. 

Fig. \ref{rgflow}. RG flow corresponding to the equations
(\ref{th19}), for $d=2$
($\epsilon=0$), with $\gamma\equiv \gamma_t$.
$\gamma/(1+\gamma)$ is plotted as a function of $t/(1+t)$ to 
transform the interval $[0,+\infty)$ into the interval $[0,1]$
on both axes.
Observe that $\gamma_t$ always scales to strong coupling, i.e.,
$\gamma/(1+\gamma) \to 1$.

Fig. \ref{rgflow3}. RG flow corresponding to 
the equations (\ref{th19}), for $d=3$
($\epsilon=1$). The notations are the same as in Fig. \ref{rgflow}.
Observe the two possible asymptotic behaviors: on the weak-disorder
side (small $t$), $\gamma$ scales to a finite value, whereas a
two-dimensional-like behavior, with $\gamma\to\infty$, is found
on the strong-disorder side (large $t$). The separatrix is marked by
a thicker line.
\newpage


~\newpage

~\newpage

~\newpage

~\newpage

~\newpage

~\newpage

~\newpage

~\newpage

~\newpage

~\newpage

\end{document}